\documentclass{nature}

\usepackage{epsfig}
\usepackage{color}
\newcommand{\gtrsim}{\lower.7ex\hbox{$\;\stackrel{\textstyle>}{\sim}\;$}}
\newcommand{\lesssim}{\lower.7ex\hbox{$\;\stackrel{\textstyle<}{\sim}\;$}}

\title{Sub-millimetre galaxies reside in dark matter halos with masses greater than $\mathbf{3\times 10^{11}}$ M$\mathbf{_{\odot}}$}

\author{A.~Amblard$^{1}$,
A.~Cooray$^{1,2}$,
P.~Serra$^{1}$,
B.~Altieri$^{3}$,
V.~Arumugam$^{4}$,
H.~Aussel$^{5}$,
A.~Blain$^{2}$,
J.~Bock$^{2,6}$,
A.~Boselli$^{7}$,
V.~Buat$^{7}$,
N.~Castro-Rodr{\'\i}guez$^{8,9}$,
A.~Cava$^{8,9}$,
P.~Chanial$^{10}$,
E.~Chapin$^{11}$,
D.L.~Clements$^{10}$,
A.~Conley$^{12}$,
L.~Conversi$^{3}$,
C.D.~Dowell$^{2,6}$,
E.~Dwek$^{13}$,
S.~Eales$^{14}$,
D.~Elbaz$^{5}$,
D.~Farrah$^{15}$,
A.~Franceschini$^{16}$,
W.~Gear$^{14}$,
J.~Glenn$^{12}$,
M.~Griffin$^{14}$,
M.~Halpern$^{11}$,
E.~Hatziminaoglou$^{17}$,
E.~Ibar$^{18}$,
K.~Isaak$^{14}$,
R.J.~Ivison$^{18,4}$,
A.A.~Khostovan$^{1}$,
G.~Lagache$^{19}$,
L.~Levenson$^{2,6}$,
N.~Lu$^{2,20}$,
S.~Madden$^{5}$,
B.~Maffei$^{21}$,
G.~Mainetti$^{16}$,
L.~Marchetti$^{16}$,
G.~Marsden$^{11}$,
K.~Mitchell-Wynne$^{1}$,
H.T.~Nguyen$^{6,2}$,
B.~O'Halloran$^{10}$,
S.J.~Oliver$^{15}$,
A.~Omont$^{22}$,
M.J.~Page$^{23}$,
P.~Panuzzo$^{5}$,
A.~Papageorgiou$^{14}$,
C.P.~Pearson$^{23,24}$,
I.~P{\'e}rez-Fournon$^{8,9}$,
M.~Pohlen$^{14}$,
N.~Rangwala$^{12}$,
I.G.~Roseboom$^{15}$,
M.~Rowan-Robinson$^{10}$,
M.~S\'anchez Portal$^{3}$,
B.~Schulz$^{2,20}$,
Douglas~Scott$^{11}$,
N.~Seymour$^{23}$,
D.L.~Shupe$^{2,20}$,
A.J.~Smith$^{15}$,
J.A.~Stevens$^{25}$,
M.~Symeonidis$^{23}$,
M.~Trichas$^{10}$,
K.~Tugwell$^{26}$,
M.~Vaccari$^{16}$,
E.~Valiante$^{11}$,
I.~Valtchanov$^{3}$,
J.~D.~Vieira$^{2}$,
L.~Vigroux$^{22}$,
L.~Wang$^{15}$,
R.~Ward$^{15}$,
G.~Wright$^{18}$,
C.K.~Xu$^{2,20}$, \&
M.~Zemcov$^{2,6}$}

\begin{document}

\maketitle

\begin{affiliations}
\item Dept. of Physics \& Astronomy, University of California, Irvine, CA 92697, USA
\item California Institute of Technology, 1200 E. California Blvd., Pasadena, CA 91125, USA
\item Herschel Science Centre, European Space Astronomy Centre, Villanueva de la Ca\~nada, 28691 Madrid, Spain
\item Institute for Astronomy, University of Edinburgh, Royal Observatory, Blackford Hill, Edinburgh EH9 3HJ, UK
\item Laboratoire AIM-Paris-Saclay, CEA/DSM/Irfu - CNRS - Universit\'e Paris Diderot, CE-Saclay, pt courrier 131, F-91191 Gif-sur-Yvette, France
\item Jet Propulsion Laboratory, 4800 Oak Grove Drive, Pasadena, CA 91109, USA
\item Laboratoire d'Astrophysique de Marseille, OAMP, Universit\'e Aix-marseille, CNRS, 38 rue Fr\'ed\'eric Joliot-Curie, 13388 Marseille cedex 13, France
\item Instituto de Astrof{\'\i}sica de Canarias (IAC), E-38200 La Laguna, Tenerife, Spain
\item Departamento de Astrof{\'\i}sica, Universidad de La Laguna (ULL), E-38205 La Laguna, Tenerife, Spain
\item Astrophysics Group, Imperial College London, Blackett Laboratory, Prince Consort Road, London SW7 2AZ, UK
\item Department of Physics \& Astronomy, University of British Columbia, 6224 Agricultural Road, Vancouver, BC V6T~1Z1, Canada
\item Dept. of Astrophysical and Planetary Sciences, CASA 389-UCB, University of Colorado, Boulder, CO 80309, USA
\item Observational  Cosmology Lab, Code 665, NASA Goddard Space Flight  Center, Greenbelt, MD 20771, USA
\item Cardiff School of Physics and Astronomy, Cardiff University, Queens Buildings, The Parade, Cardiff CF24 3AA, UK
\item Astronomy Centre, Dept. of Physics \& Astronomy, University of Sussex, Brighton BN1 9QH, UK
\item Dipartimento di Astronomia, Universit\`{a} di Padova, vicolo Osservatorio, 3, 35122 Padova, Italy
\item ESO, Karl-Schwarzschild-Str. 2, 85748 Garching bei M\"unchen, Germany
\item UK Astronomy Technology Centre, Royal Observatory, Blackford Hill, Edinburgh EH9 3HJ, UK
\item Institut d'Astrophysique Spatiale (IAS), b\^atiment 121, Universit\'e Paris-Sud 11 and CNRS (UMR 8617), 91405 Orsay, France
\item Infrared Processing and Analysis Center, MS 100-22, California Institute of Technology, JPL, Pasadena, CA 91125, USA
\item School of Physics and Astronomy, The University of Manchester, Alan Turing Building, Oxford Road, Manchester M13 9PL, UK
\item Institut d'Astrophysique de Paris, UMR 7095, CNRS, UPMC Univ. Paris 06, 98bis boulevard Arago, F-75014 Paris, France
%\item Mullard Space Science Laboratory, University College London, Holmbury St. Mary, Dorking, Surrey RH5 6NT, UK
\item Space Science \& Technology Department, Rutherford Appleton Laboratory, Chilton, Didcot, Oxfordshire OX11 0QX, UK
\item Institute for Space Imaging Science, University of Lethbridge, Lethbridge, Alberta, T1K 3M4, Canada
\item Centre for Astrophysics Research, University of Hertfordshire, College Lane, Hatfield, Hertfordshire AL10 9AB, UK
\item Mullard Space Science Laboratory, University College London, Holmbury St. Mary, Dorking, Surrey RH5 6NT, UK
\end{affiliations}

\begin{abstract}
The extragalactic background light at far-infrared wavelengths\cite{Puget1996,Fixsen1998,Dwek1998}
originates from optically-faint, dusty, star-forming galaxies in the universe with star-formation rates at the level of a few hundred solar masses per year\cite{Hughes98}.
Due to the relatively poor spatial resolution of far-infrared telescopes\cite{Nguyen2010,Hauser2001}, the faint sub-millimetre galaxies are challenging to study individually.
Instead, their average properties can be studied using statistics such as the angular power spectrum of the background intensity variations\cite{Amblard2007,Haiman2000,Knox2001,Negrello2007}.  
A previous attempt\cite{Viero2009}   at measuring this power spectrum resulted in the suggestion that
the clustering amplitude is below the level computed with a simple ansatz based on a halo model\cite{Cooray2002}.
Here we report a clear detection of the  excess clustering over the linear prediction at arcminute 
angular scales in the power spectrum of brightness
fluctuations at 250, 350, and 500 $\mu$m. From this excess,  we find that sub-millimetre galaxies are located in dark matter halos with a
minimum mass of log[M$_{\rm \mathbf{min}}$/M$\mathbf{_{\odot}}$] $\mathbf{=11.5^{+0.7}_{-0.2}}$ at 350 $\mathbf{\mu}$m. 
This minimum dark matter halo mass  corresponds to the most efficient mass scale for star formation in the universe\cite{Bouche2010},
and is lower than that predicted by semi-analytical models for galaxy formation\cite{Gonzalez2010}. 
\end{abstract}

Despite recent successes in attributing most of the extragalactic background light at sub-millimetre wavelengths
to known galaxy populations through stacking analyses\cite{Dole2006,Devlin2009,Marsden2009}, we
have not individually detected the faint galaxies that are responsible for more than 85\% of the total extragalactic intensity at these wavelengths\cite{Oliver2010}.
The faint 
star-forming galaxies are expected to trace the large-scale structure of the Universe, especially in models
where galaxy formation and evolution is closely connected to dark matter halos.
While not individually detected in low resolution observations, the clustering of 
galaxies is expected to  leave a distinct signature in the total intensity variations 
at sub-millimetre wavelengths. The amplitude of the  power spectrum of
intensity variations as a function of the angular scale provides details on 
the redshift distribution and the dark matter halo mass scale of dusty, star-forming
galaxies in the universe\cite{Amblard2007}.

For this analysis, we used data from the Herschel Multi-tiered Extra-galactic survey (HerMES\cite{Oliver2010}), 
taken with the Spectral and Photometric Imaging Receiver (SPIRE\cite{Griffin2010}) onboard the {\it Herschel} 
Space Observatory\cite{Pilbratt2010}, during the {\it Science Demonstration Phase} (SDP) of  {\it Herschel}.
The data are composed of a wide 218$'$ by 218$'$ area in the Lockman Hole complemented by a narrow, but very 
deep (30 repeated scans), map of the Great Observatories Origins Deep Survey (GOODS) North field covering 30$'$ by 30$'$.
These fields have been very well studied at other wavelengths and they are known to have a low Galactic 
dust density, making it easier to distinguish the extragalactic component we wish to study.
The observing time to complete each of the two fields  was about 13.5 hours, observing simultaneously at 250, 350, and 
500 $\mu$m. 

To limit the influence of a few bright galaxies on the measurement of the power spectrum,
we removed galaxies brighter than 50~mJy in all three passbands by masking pixels in our maps with
values larger than  50~mJy/beam as well as the neighboring pixels.
We use the cross-power spectrum of two sub-maps as our estimate of the sky power spectrum
to remove the contribution from the instrumental noise and alleviate potential systematic effects.
We correct the raw cross-power spectra for the effects of the angular response function of
the instrument and the transfer function of the map-making process. The angular response is established from 
a different set of SPIRE observations targetting Neptune, a strong point-like source for SPIRE, and involving a fine sampling of the beam 
with a  total of 700 scans\cite{Swinyard2010}.
The effects of filtering of the time-ordered data and of the map pixelization are captured  with a large set of sky simulations.
To estimate our uncertainties, we propagate the errors from the beam measurement, while the simulations
provide us with the instrumental and sky variance. The quadratic sum of these errors constitutes our error estimate.

The measured angular power spectrum (c.f. Figure~1a)
contains  contributions from spatial variations in Galactic dust clouds, or cirrus, brightness at large angular scales,
the clustering of galaxies  at intermediate angular scales, and a white-noise component at small angular 
scales arising from the Poisson behavior of the faint galaxies\cite{Knox2001,Amblard2007}. 
The cirrus signal in our Lockman-SWIRE field is taken from existing measurements in the same field 
with {\it IRAS} 100 $\mu$m and MIPS\cite{Lagache07} and extend this spectrum from 100 $\mu$m to SPIRE wavelengths
 using the spectral dependence of a Galactic dust  model\cite{FDS}. We remove 
this cirrus power spectrum from our measurements and account for 
the uncertainty of the cirrus power spectrum by adding its error in quadrature to errors of our power spectrum points.

The Poisson behaviour of sources lead to an additional term to the angular power spectrum that is scale independent.
The clustering component we measure is the excess of clustered background fluctuations above this shot-noise level.
As the confusion noise is at the level of 6 mJy at SPIRE wavelenghts\cite{Nguyen2010}, with 
fluctuations in the brightness of the background we are probing the
clustering of faint galaxies with fluxes at the level of a few mJy at 350 $\mu$m.
To extract astrophysical information on faint galaxies from the clustering power spectrum we make use of 
the  halo model\cite{Cooray2002}. This phenomenological approach connects the spatial distribution of galaxies in the universe to that of the dark matter.
To model sub-millimetre galaxies in dark matter halos, we take a halo occupation distribution describing the number of galaxies as a function of the halo mass $M$ 
of $N_{\rm gal}(M)=1+(M/M_1)^\alpha$ when $M > M_{\rm min}$ above the minimum mass scale $M_{\rm min}$,  $M_1$ 
is the mass scale at which more than one galaxy is present in a single dark matter halo, taken to be between 10 to 25 times $M_{\rm min}$,
and $\alpha$ is the power-law scaling of satellite galaxies with halo mass. The halo model involves two parts with clustering of galaxies within
halos, the one-halo term,  and clustering of galaxies between halos, the two-halo term. 
While with the two-halo term alone parameters related to the occupation number are degenrate with each other and the bias factor or the number density of galaxies,
with clustering in the one-halo part of the power spectrum aslo included the parameter degeneracies are broken and $M_{\rm min}$ can  be determined more accurately\cite{Cooray2002}.

At scales of a few arcminutes and above  we measure a clustering excess arising from the one-halo term  and above the two-halo term tracing the
linear density field power spectrum scaled by galaxy bias (c.f. Figure~1b). The one-halo term arises when more than one far-infrared 
galaxy occupies the same halo. While a hint of this one-halo term  was previously seen in the clustering of
the bright ($>30$ mJy) sub-millimetre galaxies\cite{Cooray2010}, evidence for such clustering was not present for bright galaxies
in a different {\it Herschel} dataset\cite{Maddox2010}.  To describe the power spectrum of the intensity fluctuations we also need  a prescription for the redshift evolution of the source intensity.
While models exist in the literature\cite{Lagache2003,Valiante2009}, our data are 
of sufficient quality that we can directly constrain the redshift evolution of the source intensity from our
 measurements and we constrain its value in four redshift bins in the range 0 to 4.0. The halo model parameters, the source intensity parameters, and the shot-noise amplitude
are jointly estimated with Markov-Chain Monte-Carlo fits to the power spectrum measurements. We impose an additional prior on our parameter estimates that the redshift integrated source intensity from our model fits,
including the fractional contribution from bright sources that we have masked,
be within the 68\% range of the known background light intensity at each of the three wavebands\cite{Fixsen1998}. We combine the estimates of the source intensity evolution 
at three wavebands of 250, 350 and 500 $\mu$m and in four redshift bins to measure the bolometric luminosity density between 8 and 1100 $\mu$m as a function of the redshift (c.f. Figure 2).
 We find that the far-infrared luminosity density continues to be significant out to a redshift of 4 and is at least a factor  of 10 larger than
the luminosity density of individually detected sub-mm sources with flux densities above 30 mJy alone\cite{Oliver2010}. 

Using the halo model fits, we estimate the mininum dark matter mass scale for dusty star-forming galaxies at the peak of the star formation history of the universe 
to be $\log_{10}M_{\rm min}/{\rm M}_{\odot}=11.5^{+0.7}_{-0.2}$ at 350 $\mu$m with a bias factor for the galaxies of $2.4^{+1.0}_{-0.2}$.
The minimum halo masses   $\log_{10}M_{\rm min}/{\rm M}_{\odot}$ at 250 and 500 $\mu$m are $11.1^{+1.0}_{-0.6}$ and  $11.8^{+0.4}_{-0.3}$, respectively. The corresponding bias factors for the galaxies
are  $2.0^{+0.9}_{-0.1}$ and  $2.8^{+0.4}_{-0.5}$ at 250 and 500 $\mu$m, respectively. The differences in the minimum halo masses and the bias factors between the
three wavelengths are likely due  a combination of effects including overall calibration uncertainties, 
the fact that at longer wavelengths we may be probing colder dust than at shorter wavelengths, and due to differences in the prior assumption on the total background intensity. 
In future, numerical models on the distribution of sub-millimetre galaxies
will become useful to properly understand some of these subtle differences. Averaging over the three wavelengths
the  minimum halo mass for sub-millimetre galaxies is at the level of $3 \times 10^{11}$ M$_{\odot}$, with an overall statistical
uncertainty of roughly $\pm 0.4$ in the logarithm of the minimum halo mass.

Based on a variety of  observed scaling relations such as the one between stellar mass and circular velocity, the dark matter halo mass scale for efficient star-formation 
has been  indirectly inferred to be about 10$^{11}$ M$_{\odot}$\cite{Bouche2010}. 
As the sub-millimetre galaxies are the most active star-forming galaxies in the universe, it is likely that the minimum halo mass scale 
for such galaxies that we determine from brightness fluctuations corresponds to the preferred mass scale of active star-formation in the universe. 
In dark matter halos below this mass, star-formation is expected to be inefficient due to photo-ionization feedback\cite{Bouche2010}.
The underlying  astrophysics needed to explain the numerical value we find are still missing in galaxy formation theories, since
existing semi-analytical models predict a mass scale for faint sub-millimetre galaxies that are roughly ten times larger\cite{Gonzalez2010}.  
We provide the strongest evidence for a minimum mass scale for active 
star-formaing galaxies in the universe by stuyding the background intensity variations generated by those galaxies in our sky maps.
Our direct estimate of the minium dark matter halo mass provides a critical value needed to improve theoretical models 
of sub-millimeter galaxies and the overall galaxy formation and evolution.

\begin{addendum}
 \item 
SPIRE has been developed by a consortium of institutes led by Cardiff University (UK) and including Univ. Lethbridge (Canada); NAOC (China); CEA, LAM (France); 
IFSI, Univ. Padua (Italy); IAC (Spain); Stockholm Observatory (Sweden); Imperial College London, RAL, UCL-MSSL, UKATC, Univ. Sussex (UK); and Caltech/JPL, IPAC, 
Univ. Colorado (USA). This development has been supported by national funding agencies: CSA (Canada); NAOC (China); CEA, CNES, CNRS (France); ASI (Italy); MCINN (Spain); 
SNSB (Sweden); STFC (UK); and NASA (USA). We thank M. Viero for useful comments. 
Amblard, Cooray, Serra, Khostovan, Mitchell-Wynne, and other US co-authors are supported by NASA funds for US participants in {\it Herschel} through an award from JPL.
The data presented in this paper are publicly available from the ESA/Herschel Science Archive (http://herschel.esac.esa.int)
under the observational identifications 1342186108, 1342186109, and 1342185536. Derived products by the HerMES collaboration, such as source catalogs,
will be released through the HeDaM Database (http://hedam.oamp.fr/HerMES).

\item[Author Contributions] This paper represents the combined work of the HerMES Collaboration, the SPIRE Instrument Team's Extragalactic Survey, and
has been extensively internally reviewed. A.C. planned the study, supervised the research work of A.A. and P.S. and wrote the draft version of this paper.
A.A. performed the power spectrum measurements
and P.S. interpreted those measurements with the halo model. All other coauthors of this paper contributed extensively and equally by their varied contributions to the SPIRE
instrument, Herschel Mission, analysis of SPIRE and HerMES data, planning of HerMES observations, scientific support of HerMES,
and by commenting on this manuscript as part of an internal review process.

 \item[Correspondence] Correspondence and requests for materials
should be addressed to A.C. (acooray@uci.edu)
 \item[Competing interests statement] The authors declare no    competing interests.
 \item[Supplementary information] accompanies this paper.

\end{addendum}

\clearpage
\newpage

%%
%% TABLES
%%
%% If there are any tables, put them here.
%%

\begin{figure}
\caption{{\bf The two-dimensional power spectrum of the Herschel map}.
{\it (a):} 
The total power spectrum $P(k)$ at 350 $\mu$m as a function of the wavenumber $k$ in arcmin$^{-1}$. 
The error bars are 68\% confidence level uncertainties.  The shaded region is the cirrus signal in our Lockman-SWIRE field.
To describe the total power spectrum, we take a power-law with
$P(k)=A(k/k_1)^n+P_{\rm SN}$ where $k_1$ is fixed at $0.1$ 
arcmin$^{-1}$ and $P_{\rm SN}$ is the shot-noise amplitude.  At 350 $\mu$m, we find $A=(5.79 \pm 0.26)\times10^3$ Jy$^2$/sr and $n=-1.28 \pm 0.07$, shown
by the red-line. The light blue line is the best-fit shot-noise amplitude with a value of $4600 \pm 70$ Jy$^2$/sr, in agreement with the value of $4500 \pm 220$ Jy$^2$/sr predicted by best determined source counts\cite{Glenn2010}. The shot-noise errors include the 15\% uncertainty in the SPIRE absolute flux calibration\cite{Swinyard2010}.
In green we show the total power spectrum combined with the mean estimate of the cirrus signal. 
{\it (b):} The angular clustering power spectrum at 350 $\mu$m as a function of the wavenumber with errors showing the 68\% confidence level uncertainties. The best-fit shot-noise value (dashed blue line) 
has been removed from the data and its uncertainty added to the overall error in quadrature. 
The green lines show the best-fit halo model with a reduced chi-square value of 1.02.
The dot-dashed line show the two-halo term and the dashed line show the one-halo term that is responsible for the clustering at small angular scales. 
Data in the lowest wavenumber bins contaminated by the Galactic cirrus have been omitted. For comparison we also show a previous  measurement of the power spectrum of brightness fluctuations at 350 $\mu$m with BLAST\cite{Viero2009}. The results related to 250 and 500 $\mu$m are summarized in the Supplementary Information.
}
\label{pkpsw}
\end{figure}

\begin{figure}
\caption{{\bf Far-infrared bolometric luminosity density (between 8 and 1110 $\mu$m) and star formation rate as a function of redshift}.
The far-infrared bolometric luminosity density was estimated using the values of the source intensity in four redshift bins between 0 and 4
derived from best-fit model fits to each of the power spectra from the three wavebands and with a prior selection of models with $\alpha >1$ for the power-law slope of the occupation number.
We have assumed a modified black-body spectrum for the spectral energy distribution of the spatially unresolved sources, with an emissivity index of 1.5 and a dust temperature
of 28$\pm 8$ K\cite{Amblard2010}.  We propagate the uncertainty
on the dust temperature as an additional error when computing errors on the luminosity density in each of the redshift bins.  
The vertical  bars show the 68\% confidence level errors propagated from errors in the source intensity evolution estimates and from the
temperature prior, while the horizontal bars indicated the redshift range of each of the bins we use for model fits. The orange and blue shaded areas represent the model
from  Lagache et al.\cite{Lagache2003} and Valiante et al.\cite{Valiante2009} with the same flux cut (S$>$ 50 mJy) and the
same temperature prior as our data. Our measurements are consistent with both these models.
The redshift evolution of the far-infrared luminosity is a measure of the star formation history of the universe\cite{Kennicutt1998}
and converting our estimate, we find a star formation rate of $<0.01$, $0.05 \pm 0.03$, $0.04 \pm 0.03$, and $0.06\pm0.04$ M$_\odot$/yr/Mpc$^3$ in each of the redshift bins of
$z < 1$, $1 < z < 2$, $2 < z <3$, and $3 < z <4$, respectively.
}
\label{dlfir}
\end{figure}

\clearpage

\newpage

\centerline{
\includegraphics[width=11cm]{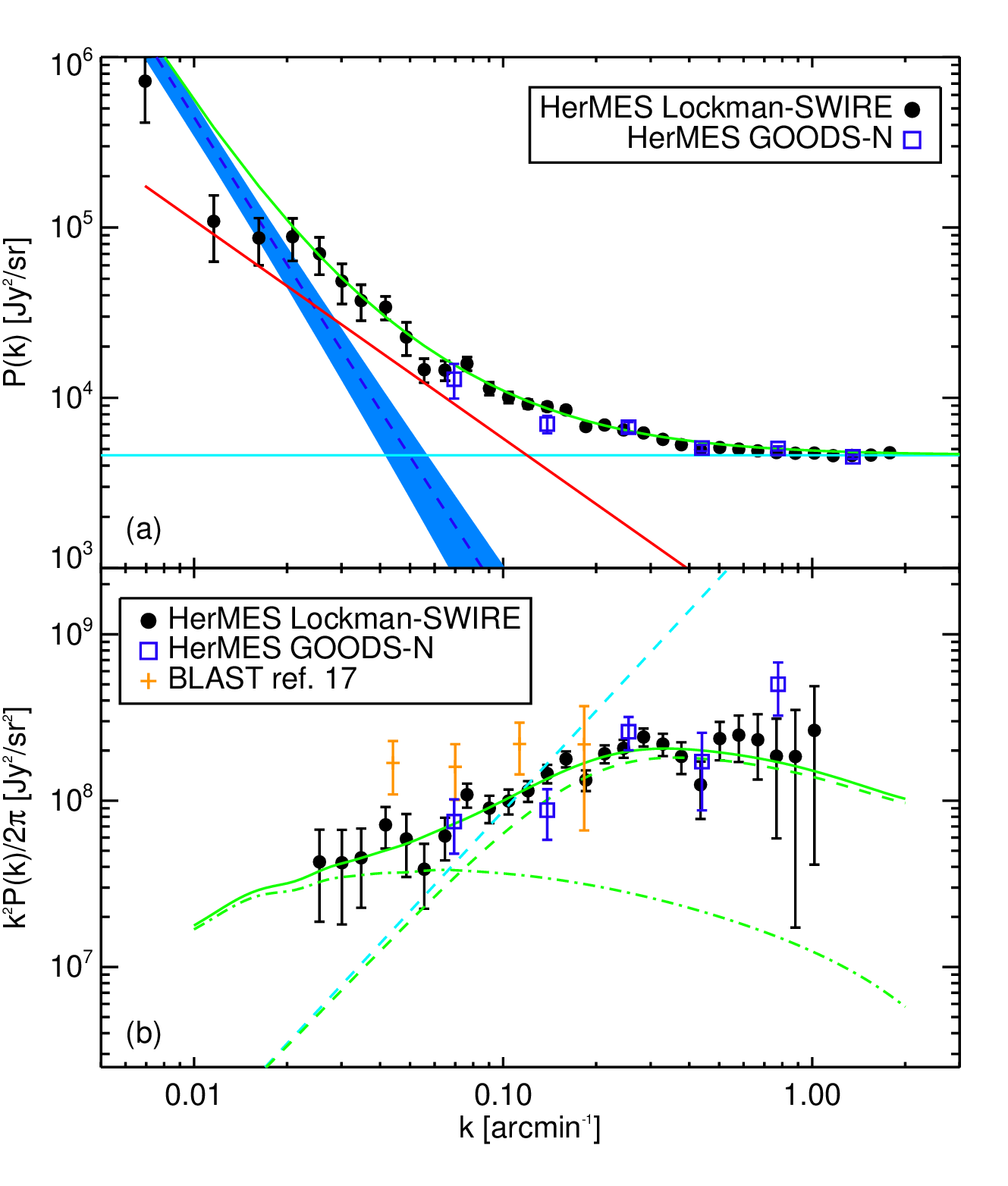}
}
%\centerline{
%\includegraphics[width=11cm]{Fig1b.ps}
%}

\newpage
\centerline{
\includegraphics[width=12cm]{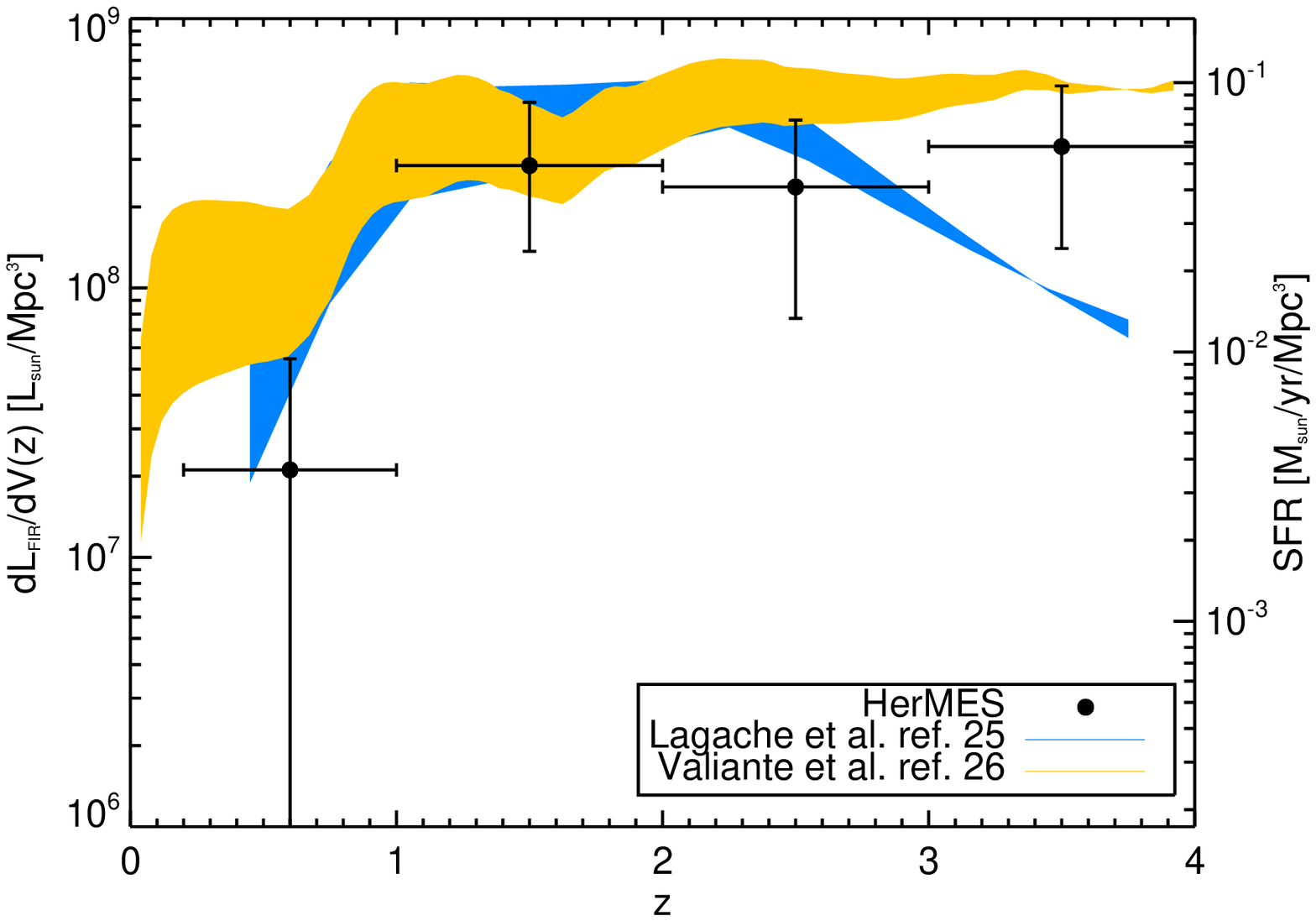}
}

\clearpage

\newpage

\setcounter{figure}{0}
%\newcounter{multibblnoresetbibitemcount}{30}
\renewcommand{\figurename}{Figure~S}

\setcounter{table}{0}
%\newcounter{multibblnoresetbibitemcount}{30}
\renewcommand{\tablename}{Table~S}

%\begin{center}
%{\bf \large Supplemental Notes: Tracing the Sources of the Cosmic Far-Infrared Background with Unresolved Fluctuations}
%\end{center}

In these Supplemental Notes to our main paper, we outline key details related to how we
estimated the angular power spectrum of {\it Herschel}-SPIRE data and its interpretation.

The data presented in the main paper are publicly available from the ESA/Herschel Science Archive (http://herschel.esac.esa.int)
under the observational identifications 1342186108, 1342186109, and 1342185536. Derived products by the HerMES collaboration, such as source catalogs,
will be released through the HeDaM Database (http://hedam.oamp.fr/HerMES).

\subsection{Data Analysis and Map making:}

The data were taken in the standard Scan-Map AOT for which the scanning speed
is 30''/s. Calibrated time-ordered data  were created using HIPE\cite{Ott2010a}
development version 2.0.905, with a fix applied to the astrometry
(included in more recent versions of the pipeline),  with newer
calibration files (SPIRE Beam Steering Mirror calibration version 2, 
flux conversion version 2.3 and temperature drift correction version 2.3.2) and 
with a median slope subtracted from each timeline. 
We removed a few percent of the data samples which were contaminated by cosmic-rays or instrumental effects and flagged by the initial pipeline.  

We convert time-ordered data to a map on the sky through an iterative baseline removal and an iterative calculation of detector weights. 
We give a summary of the map-making method here. Full details of this approach is available elsewhere\cite{Fixsen2000,Levenson2010}.

Given sky brightness $I({\bf \theta})$, the signal for a detector $d$ in a scan $s$ with a time sample $j$ can be written as
\begin{equation}
S_{dsj} = I({\bf \theta}_{dsj})+P^n_{ds}+N_{dsj} \, ,
\end{equation}
where $P^n_{ds}$ is an n-th order polynomial baseline offset for detector $d$ and scan $s$, and 
$N_{dsj}$ is the instrumental noise. The parameters of $P^n_{ds}$
are solved with an iterative  solution to the best-intensity of the sky at each step.

At each iteration $i$, we minimize the variance of the residual $V^i_{dsj}$ based on the previous map $I^{i-1}({\bf \theta}_{dsj})$ such that
\begin{equation}
V^i_{dsj} = S_{dsj}-\left[I^{i-1}({\bf \theta}_{dsj})+P_{ds}^{n,i}\right] \, .
\end{equation}
The $i$-th estimate of the sky is performed via the weighted mean of all the samples that fall into a given pixel:
\begin{equation}
I^i({\bf \theta}) = \frac{\sum_{dsj} w^i_{ds} \left(S_{dsj}-P_{ds}^{n,i}\right)}{\sum_{dsj} w^i_{ds}} \, .
\end{equation}
The weight associated with each iterative estimate is simply the inverse variance of the residual
\begin{equation}
w^i_{ds} = \frac{N_{\rm tot}}{\sum_{k=1}^{N_{\rm tot}} \left(V^i_{dsj}\right)^2} \, .
\end{equation}

\begin{figure}
\centerline{
\includegraphics[width=14cm]{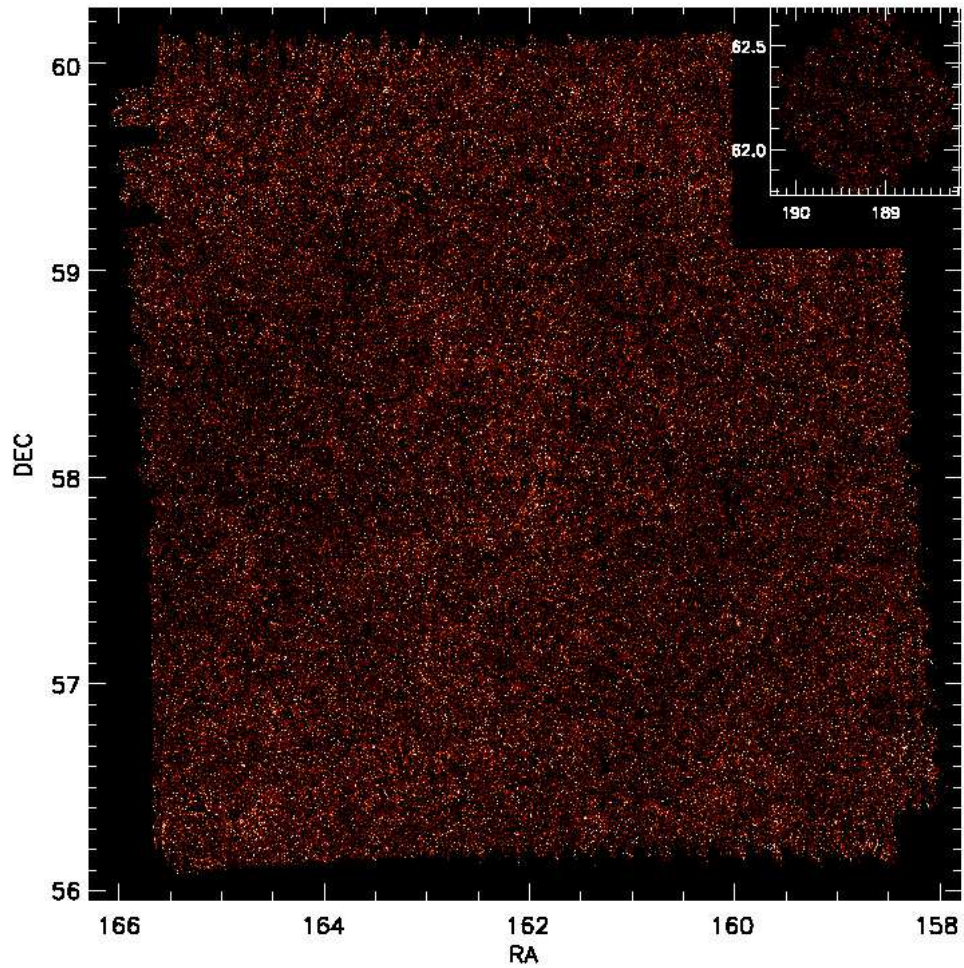}
}
\caption{
{\bf The 250 $\mu$m SPIRE maps of Lockman-SWIRE and GOODS-N (top-right inset) fields.} We have masked all galaxies above 50 mJy.
}
\label{maps}
\end{figure}

The algorithm also allows us to create a noise map by propagating detector noise as estimated by the variance of the residuals.
The maps used here make use of a first-order  polynomial with $n=1$ and 20 iterations. The gain offsets were computed from the 1st iteration while
the weights are fixed to 1.0 for the 10 iterations and are calculated from the data starting at iteration 11, to improve stability of the
algorithm. The same procedure is repeated for both Lockman-SWIRE and  GOODS-N maps and also iterative 
maps of Neptune that we use for beam measurements. The maps at 250 $\mu$m are shown in Figure~S1.

The absolute astrometry of the maps was corrected by stacking at the positions of {\it Spitzer} 
Multi-Band Imaging Photometer (MIPS) 24 $\mu$m and radio sources, finding
reasonably consistent results between the two within 0.5$''$. 
We have made an overall correction to the absolute astrometry of the order of a few arcseconds, though such small angular scale
corrections do not impact results we present here focusing between 0.5 arcminutes to 100 arcminutes.

The maps were made with pixels of size 6, 8.3, and 12 arcseconds at 250, 350, 500 $\mu$m, respectively,  corresponding to 
one third of the full-width at half-maximum (FWHM) of the SPIRE beam profiles\cite{Swinyard2010a} in the three passbands.

\subsection{Raw Power Spectra}

To compute the power spectrum in our map, we use fast Fourier transforms; however, we need to take
into account the missing and unwanted pixels. Due to our scanning strategy and some corrupted data,
a small fraction of pixels are not defined on the map that we use to define our Fourier transform basis.
Furthermore, we wish to remove the brightest galaxies in order to reduce the shot-noise term in the 
power spectrum, since the shot-noise term is weighted towards bright galaxies and larger shot-noise degrades
the ability to extract the clustering component of the power
spectrum. In our analysis we applied a flux cut of 
50 mJy/beam, removing 0.9, 0.7, 1.2\% of the pixels at 250, 350, 500 $\mu$m, respectively, in the Lockman-SWIRE field and
0.5, 0.2, 0.2\% in the GOODS-N field. We used the same flux cut at all
frequencies for simplicity, this 50 mJy/beam allows to remove all the
bright sources while retaining most of the pixels in each map.
The remaining number of pixels used for the fluctuation study is 
5.4$\times10^6$ , 2.9$\times10^6$, and $1.4\times10^6$ at 250, 350, 500 $\mu$m, respectively, in the
Lockman-SWIRE field and $1.9\times10^5$, $1.0\times10^5$, $4.7\times10^4$ in the GOODS-N field.

The raw power spectra are summarized in Figure~S2. Here, we show the auto spectra in the total map as well as the cross spectrum
with maps made with half of the time-ordered data in each map. The difference of the two provides us with an estimate
of the instrumental noise. At small physical scales (large $k$ values)
the noise is almost white such that $P(k)$ is a constant value. We fit a model of the form
\begin{equation}
N(k) = N_0 \left[\left(\frac{k_0}{k}\right)^2+1\right] \, ,
\end{equation}
and determine the knee scale of the noise, $k_0$, to be at about 0.15 arcmin$^{-1}$ at each of 250, 350, and 500 $\mu$m.

\begin{figure}
\centerline{
\includegraphics[width=14cm]{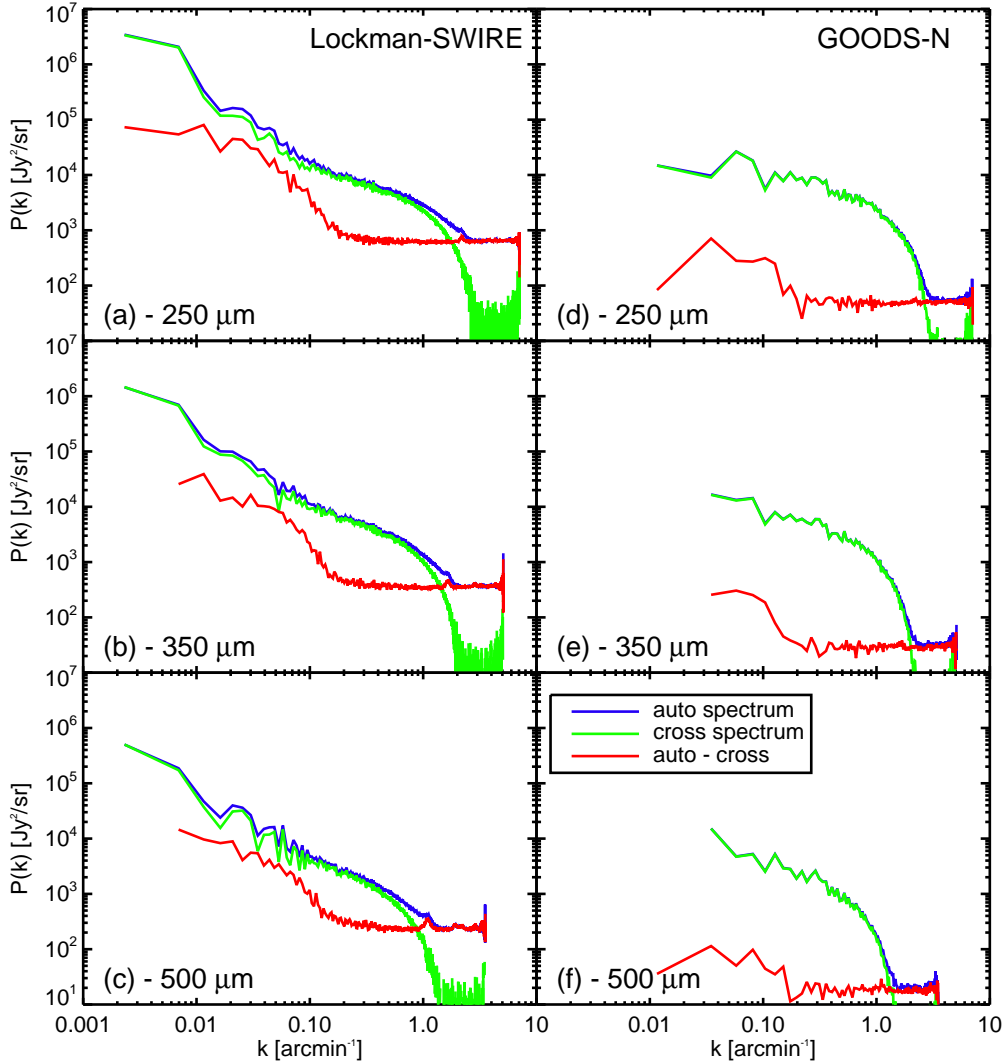}
}
\caption{{\bf Raw $P(k)$ of the Lockman-SWIRE (left) and GOODS-N (right) fields.} The panels show the three wavebands with 
 250 $\mu$m (panels {\it a} and {\it d}), 350 $\mu$m ({\it b} and {\it e}), 
and 500 $\mu$m ({\it c} and {\it f}) from top to bottom on each side, respectively.  The blue lines show
the auto-power spectra computed from maps using all the data. 
This power spectrum is a combination of sky signal and instrumental noise.
We estimate the sky signal through the cross-spectrum (green lines) of two maps after dividing the data into two halves separated in time. 
The difference of these two spectra represents an estimate of the instrumental noise power spectrum (red line).
}
\label{pkraw}
\end{figure}

\subsection{Corrected Power Spectra:}

The final power spectrum we show in this paper is corrected for a combination of effects described by 
\begin{equation}
P(k') = B(k) T(k) M_{k'k} P(k) \, ,
\label{eqn:pk}
\end{equation}
where $P(k')$ is the observed power spectrum from data in the presence of mask, $B(k)$ is the beam function, and the map making transfer
function is $T(k)$. $P(k)$ is the true sky power spectrum and is determined by inverting the above equation.

\begin{figure*}
\centerline{
\includegraphics[width=10cm]{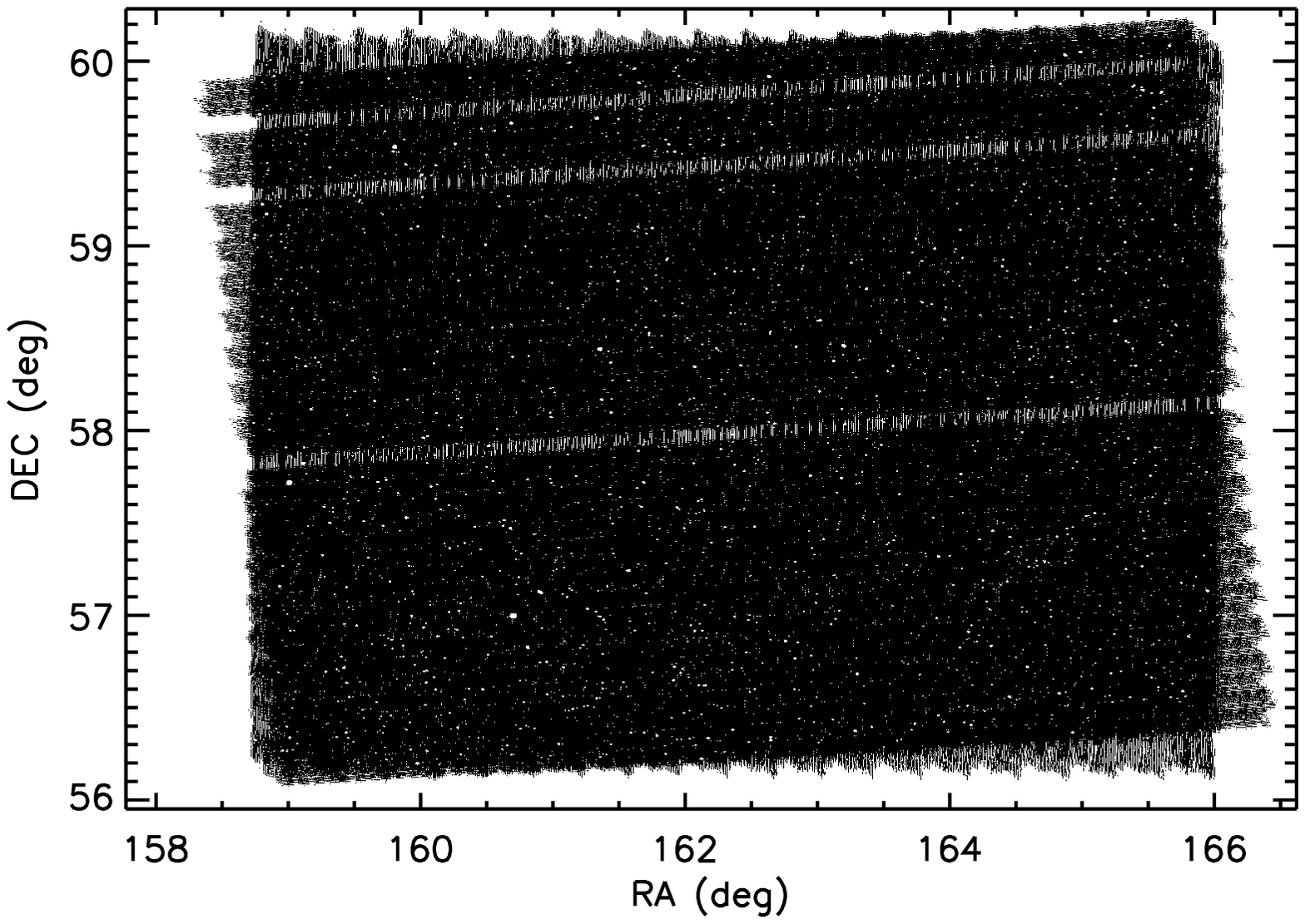}\hspace{-1cm} 
\includegraphics[width=10cm]{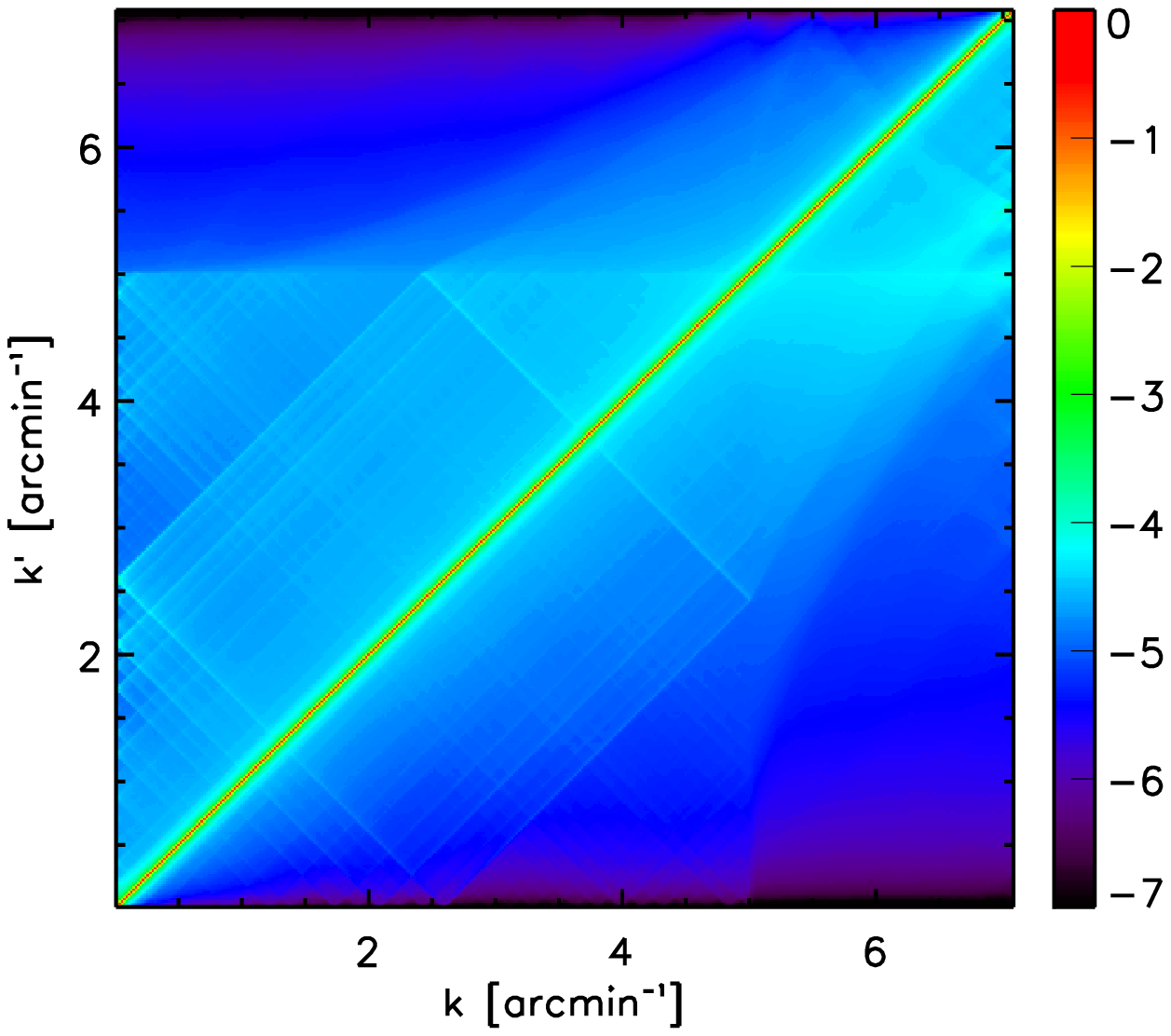}
}
\centerline{
\includegraphics[width=10cm]{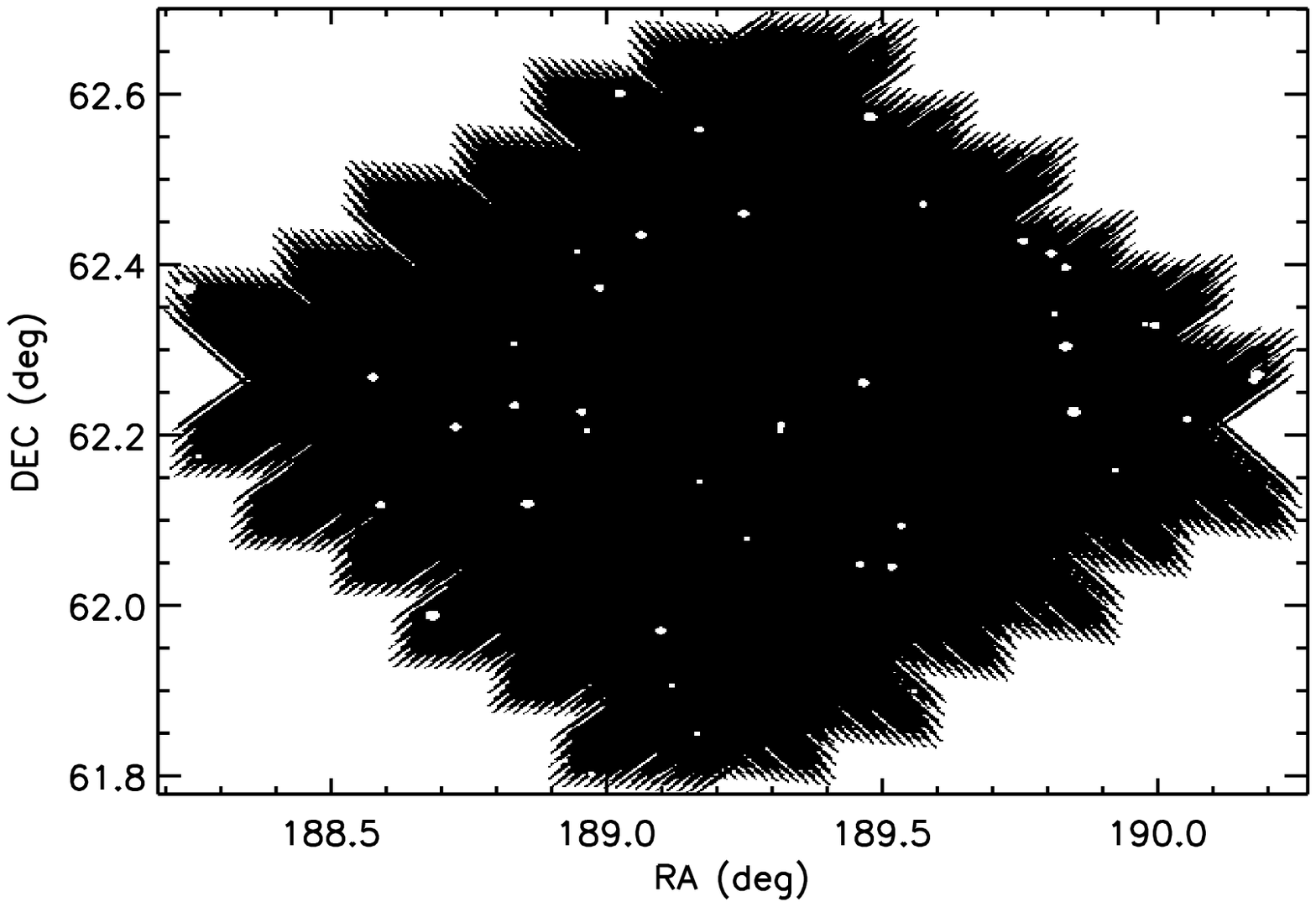}\hspace{-1cm} 
\includegraphics[width=10cm]{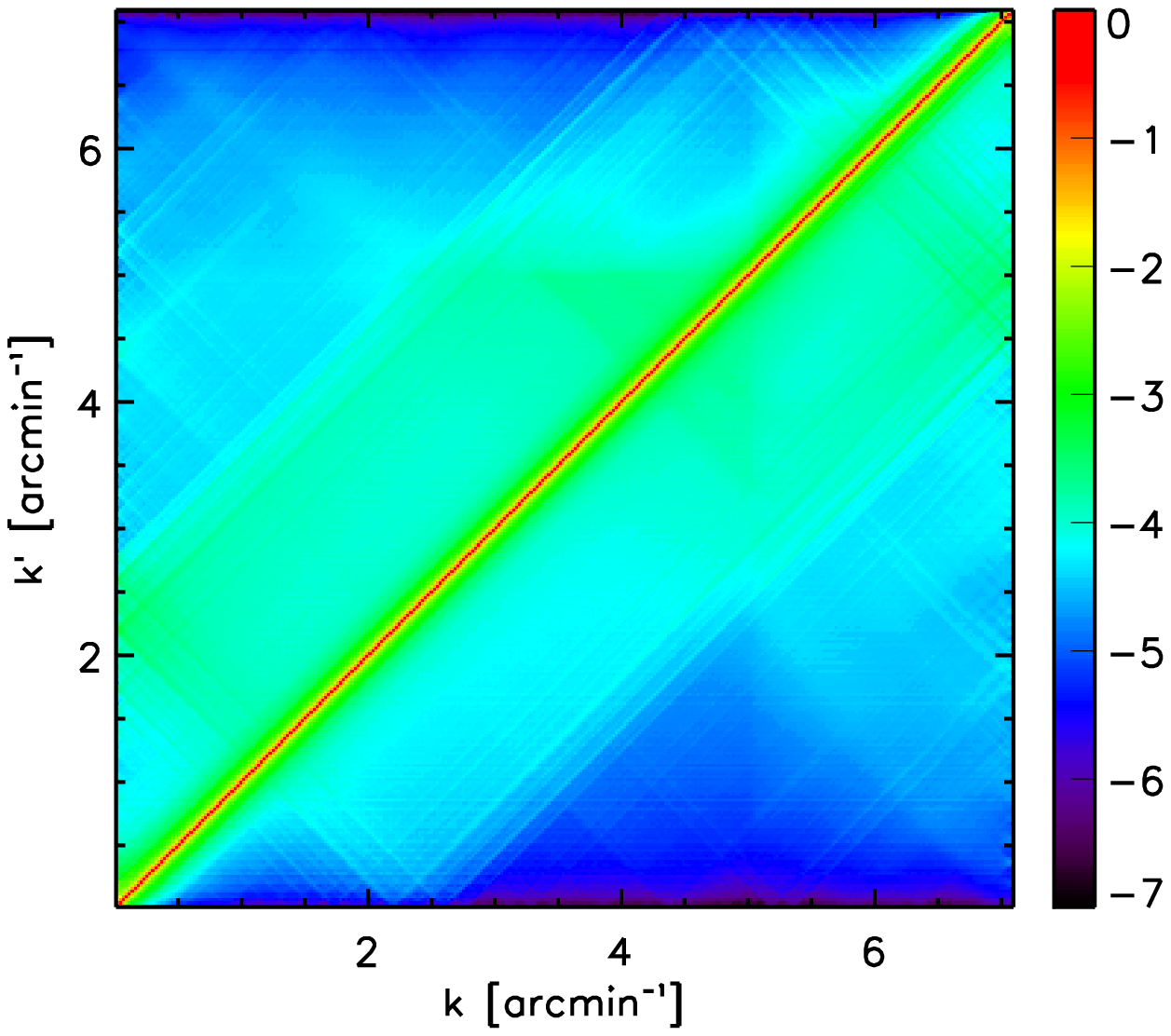}
}
\caption{{\bf Mask and the model coupling matrix.} {\it Top-left :} Mask used for the Lockman-SWIRE field. The galaxies brighter than 50 mJy masked as well as some corrupted scans.
{\it Top-right:} Coupling matrix $M_{kk'}$ (log scale) computed for this Lockman-SWIRE mask.
Bottom-left and bottom-right figures are the same things for the GOODS-N field.
}
\label{figmkk}
\end{figure*}

In the above, $M_{kk'}$ is the mode coupling matrix associated with the mask.
This can be expressed analytically in the flat sky approximation\cite{Hivon2002}  as
\begin{equation}
M_{kk'}= \sum_{\theta_k}\sum_{\theta_{k'}}{|w(k-k')|^2}/N(\theta_k) \, ,
\end{equation}
where $w(k)$ is the Fourier transform of the mask.
Figure~S\ref{figmkk} shows the mask we used and the corresponding matrix M$_{kk'}$ for each of Lockman-SWIRE and GOODS-N fields at 250 $\mu$m with sources
above 50 mJy and spurious data removed.

\begin{figure}
\hspace{-0.7cm}
\centerline{
\includegraphics[width=16cm]{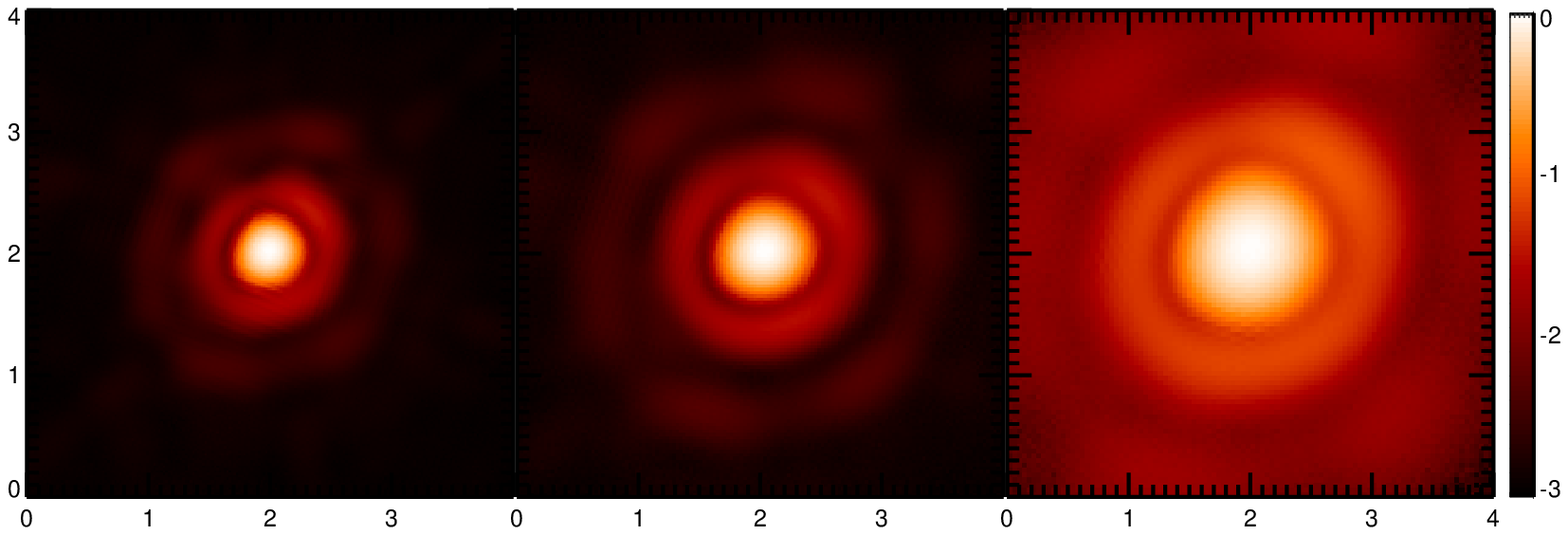}
}
\caption{{\bf Maps used for the beam function measurements.} From left to right, Neptune beam maps (log scale normalized to the maximum) at 
250, 350, and 500 $\mu$m, respectively. The x- and y-axes are labeled in arcminutes.
The maps were made with the iterative map-maker  and observations involve a total of 700 scans of Neptune.
}
\label{beammaps}
\end{figure}

\begin{figure*}
\centerline{
\includegraphics[width=6cm]{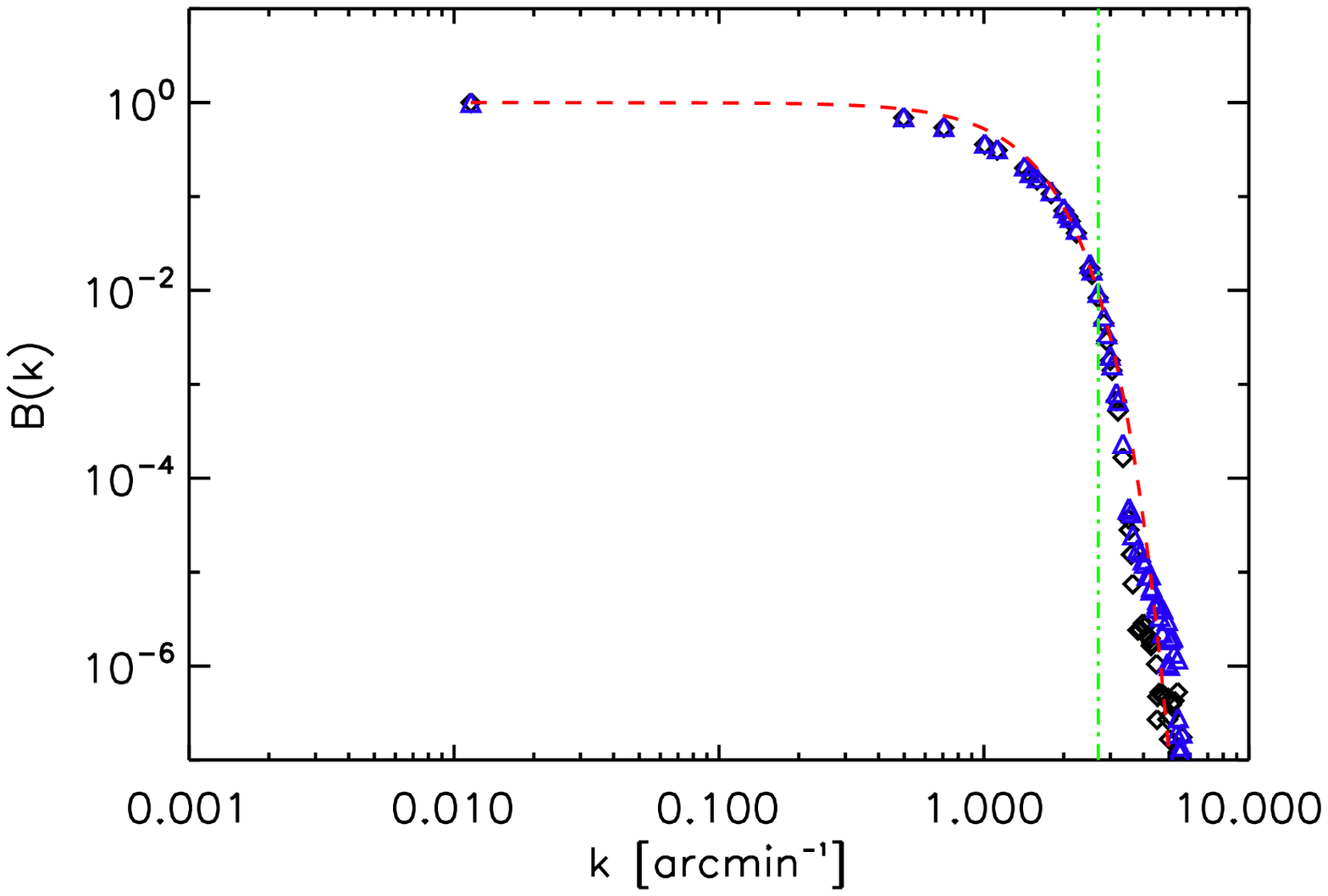}\hspace{-0.8cm} 
\includegraphics[width=6cm]{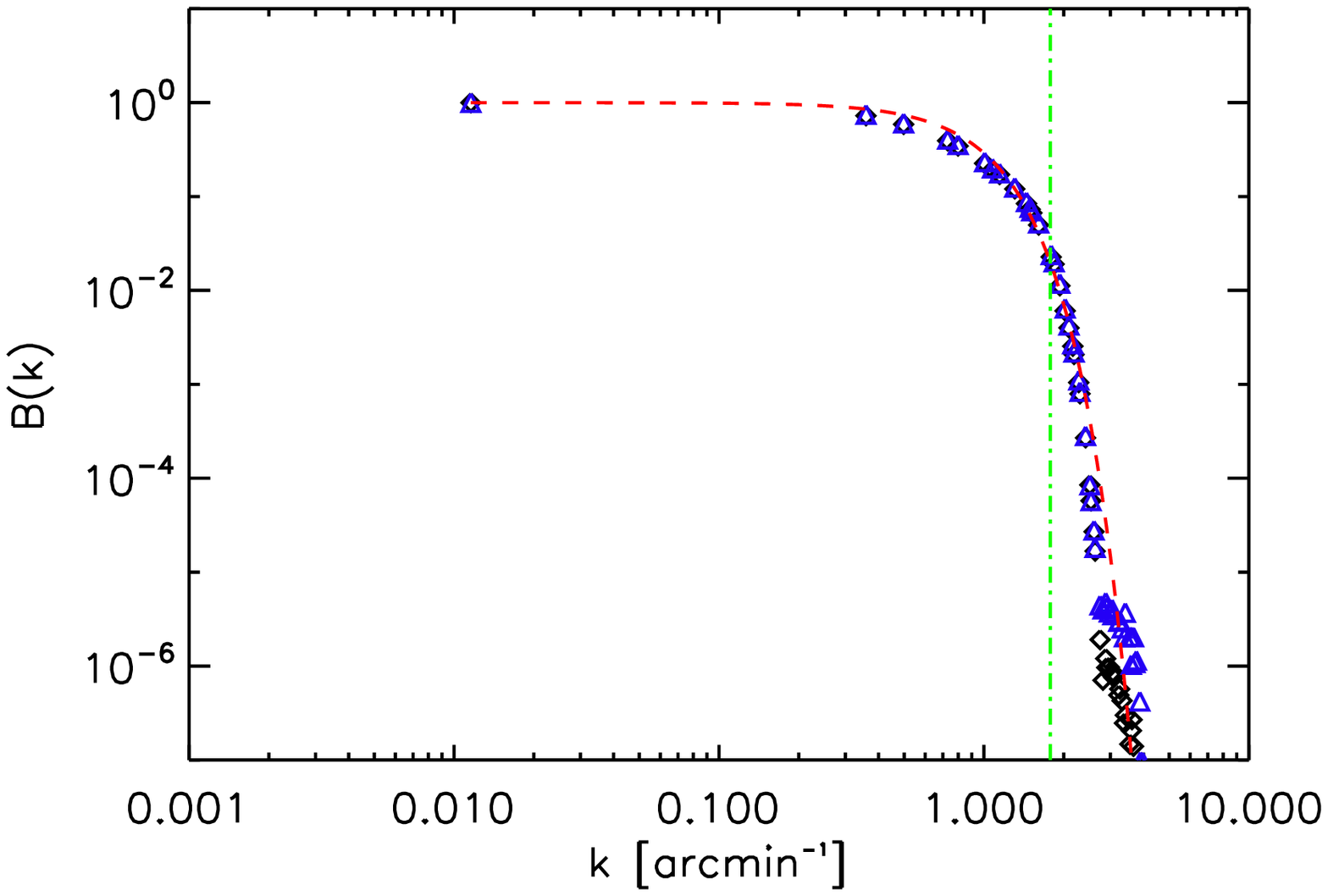}\hspace{-0.8cm}
\includegraphics[width=6cm]{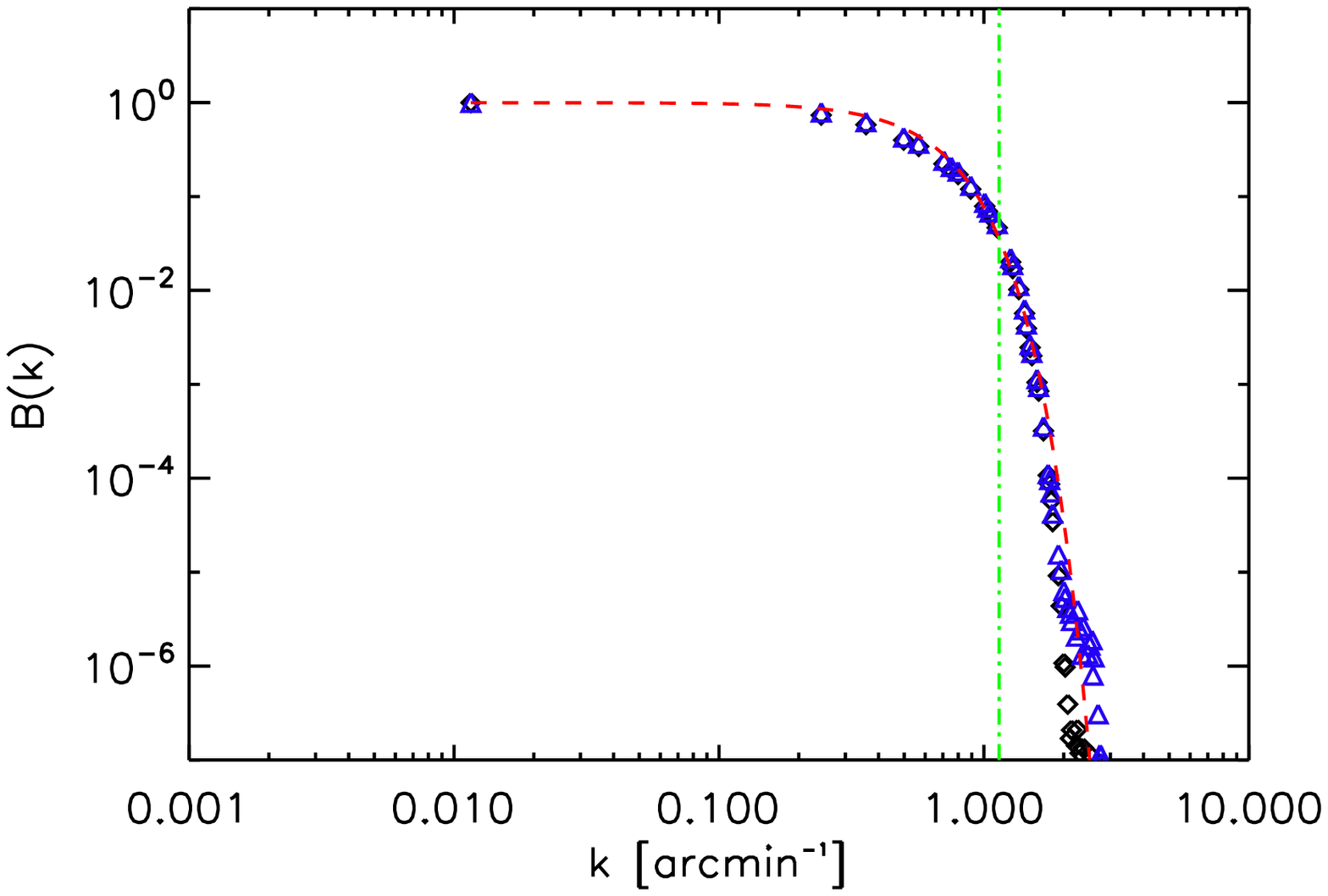}
}
\centerline{
\includegraphics[width=6cm]{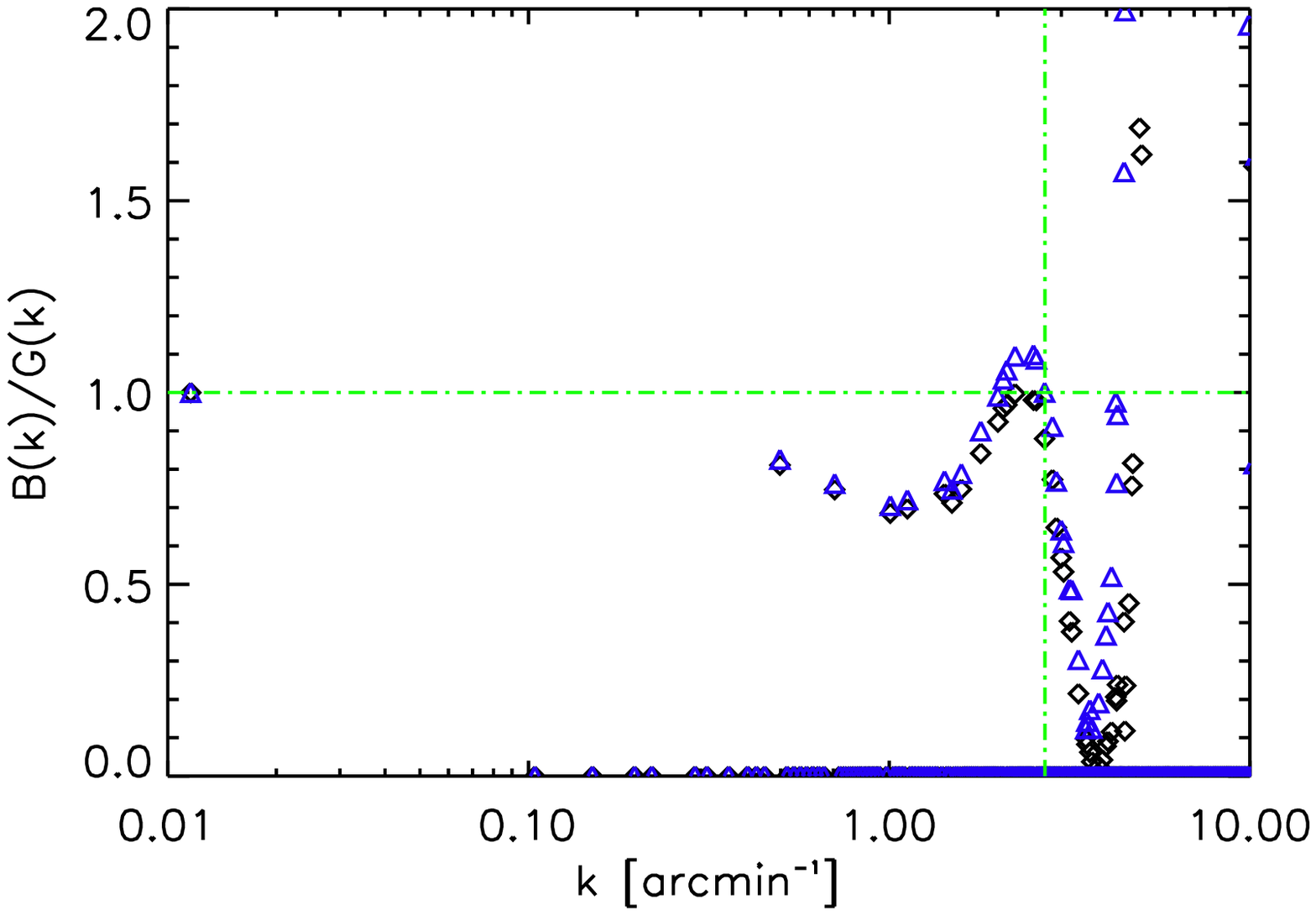}\hspace{-0.8cm} 
\includegraphics[width=6cm]{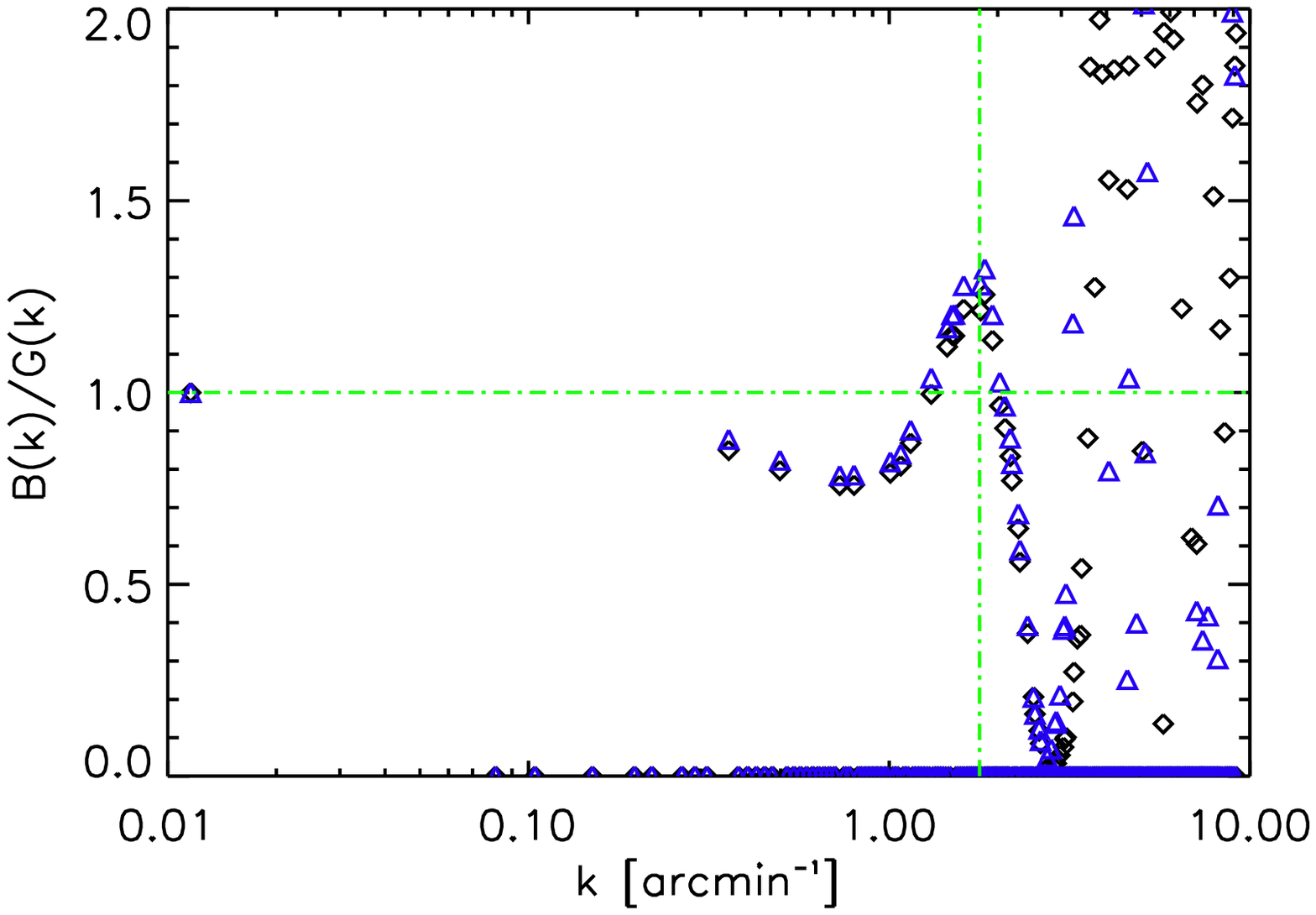}\hspace{-0.8cm}
\includegraphics[width=6cm]{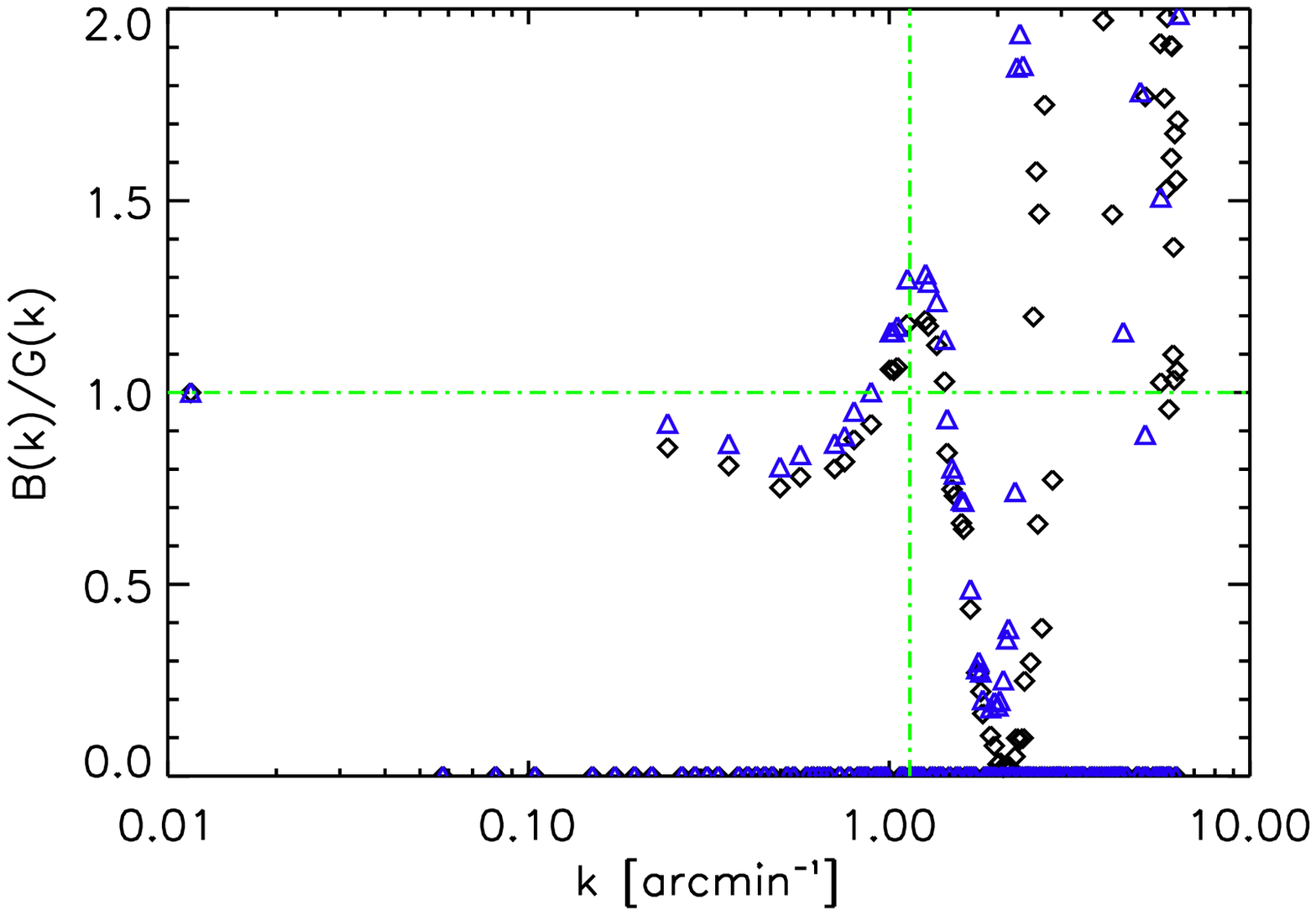}
}
\caption{{\bf Point spread function of SPIRE Instrument}. {\bf Top :} Point Spread Function Fourier space kernels for 250, 350 and 500 $\mu$m (from left to right).
The black diamonds were estimated using the Neptune ``naive'' map, the blue triangles with the Neptune  ``iterative'' map 
and the red dashed line is the Gaussian (FWHM of 18, 25 and 36 arcseconds). The vertical green
dashed dotted lines represent the maximum k out to which the data are used in this analysis.
 {\bf Bottom:} Ratio of the beam kernel measured on Neptune to the Gaussian beam approximation.
}
\label{beamfunc}
\end{figure*}

We measure the beam function $B(k)$ through observations of Neptune, involving a total of 700 scans.
Figure~S\ref{beammaps} shows the Neptune maps made at 250, 350, and 500 $\mu$m.
The Neptune data are analyzed in the same manner as the Lockman-SWIRE and GOODS-N data using the same iterative map maker. In addition we also employ 
a naive map maker available as part of HIPE\cite{Ott2010a}. 
When making these maps, we account for the relative motion of Neptune relative to the background sky
and make maps that correct for Neptune's varying position during the observations. This results in a map
where extragalactic sources are smeared. Neptune, however, is several orders of magnitude brighter
and our beam measurements primarily focus on the central region.
Figure~S\ref{beamfunc} summarizes the results related to $B(k)$ for each of the three SPIRE wavelengths.
In the same figure (bottom panels), we also compare the beam measured from Neptune to the beam described 
by a Gaussian with a FWHM
 of 18, 25, and 36$''$ at 250, 350, and 500 $\mu$m. The amplitude of $B(k)$ is thereafter interpolated in the $k$ 
modes at which we compute our fluctuation power spectra.

\begin{figure}
\hspace{-0.7cm}
\centerline{
\includegraphics[width=14cm]{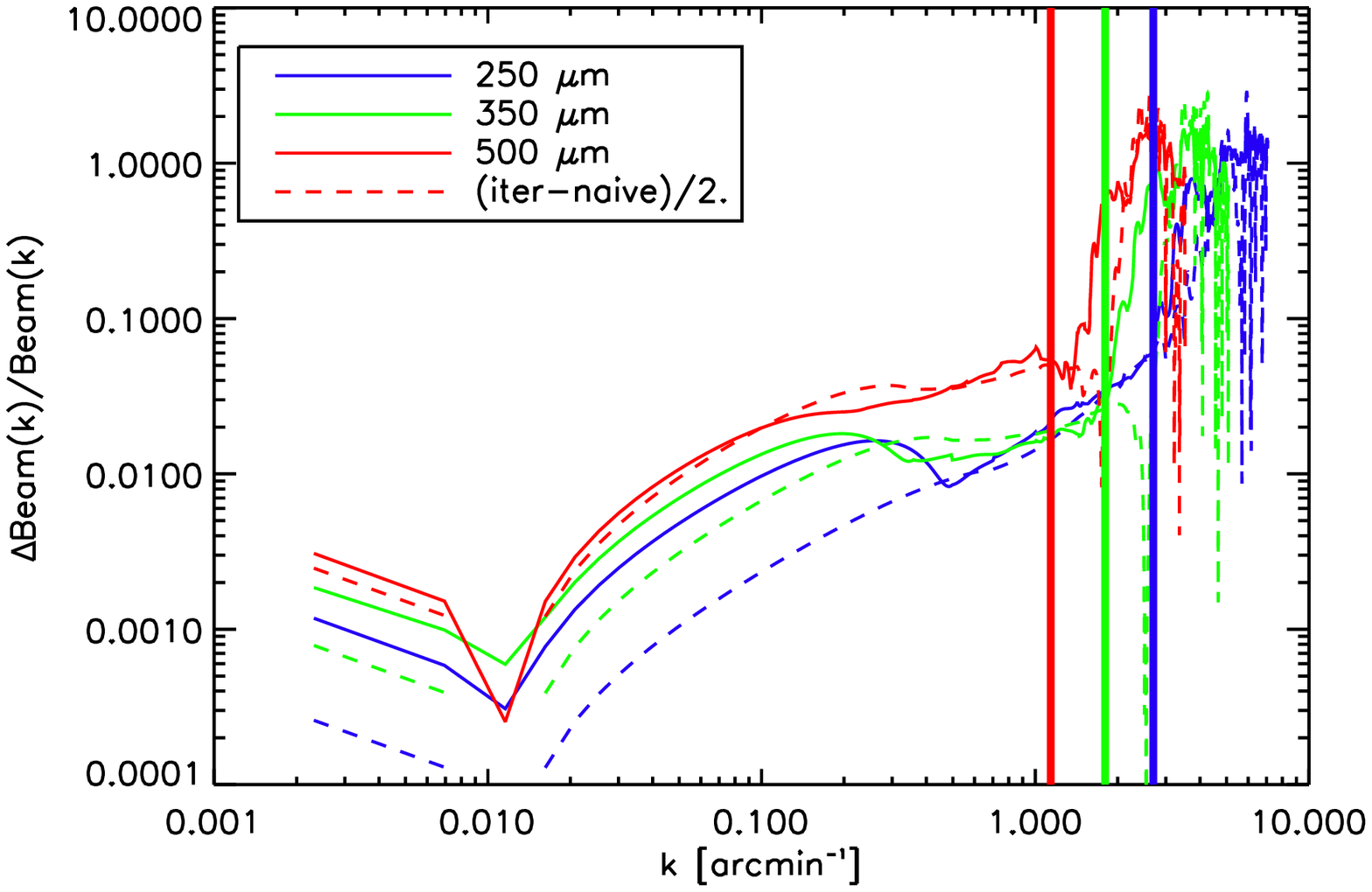}
}
\caption{{\bf Accuracy of the beam measurement.}
The beam uncertainty relative to the mean beam function used for this analysis. The total uncertainty (solid lines) is the standard deviation of the different $B(k)$ estimates using the measurements on the iterative and naive map and several interpolation methods. The dashed lines are half of the difference between one of the naive map $B(k)$ estimates and
one of the iterative map $B(k)$ estimates. The vertical lines mark the maximum $k$ value out to which we make use of the power spectrum measurements at each of 250, 350, and 500 $\mu$m.
}
\label{beamdiff}
\end{figure}

\begin{figure}
\hspace{-0.7cm}
\centerline{
\includegraphics[width=14cm]{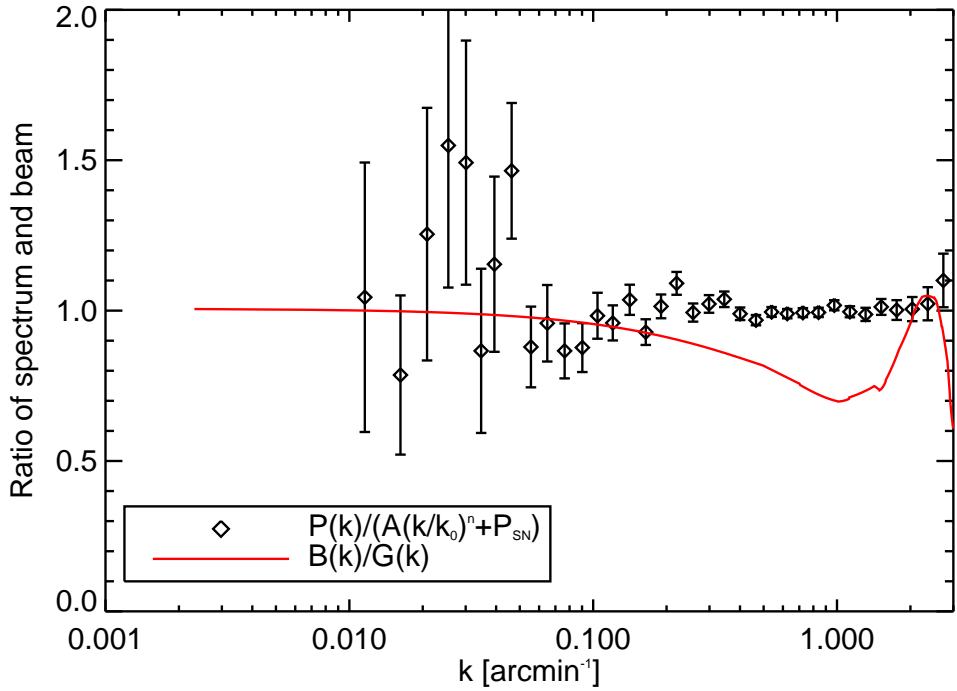}
}
\caption{{\bf Power spectrum relative to the beam function}. Comparison of the $P(k)$ estimate and the beam transfer
  function shape. The black diamonds represent our 250 $\mu$m $P(k)$
  estimate divided by the best fit power-law. The red line represents 
  our beam transfer function divided by the approximate Gaussian
  beam. The two curves have different shapes and this difference indicates that the $P(k)$
  shape does not come from the beam transfer function.
}
\label{beamcomp2pk}
\end{figure}

The uncertainty in the beam function $B(k)$ is determined by computing the standard deviation of the different
$B(k)$ estimates, using the measurements on the iterative and naive map and several different interpolation schemes.
The beam uncertainty computed this manner is slightly larger than the difference in the beams between
 two different observations of Neptune, one involving the fine scans we primarily use here and an older coarse set of Neptune scans, but with maps
made using the same map-maker.
Figure~S\ref{beamdiff} shows the overall uncertainties in the beam (solid lines) as well as the uncertainties coming from
the difference between the naive and iterative map reconstructions (dashed lines). Figure~S\ref{beamcomp2pk} compares the beam function and the power spectrum $P(k)$ at 250 $\mu$m
from the Lockman-SWIRE field showing that features in the power spectrum are not related to features in the beam function $B(k)$.

\begin{figure}
\hspace{-0.7cm}
\centerline{
\includegraphics[width=12cm]{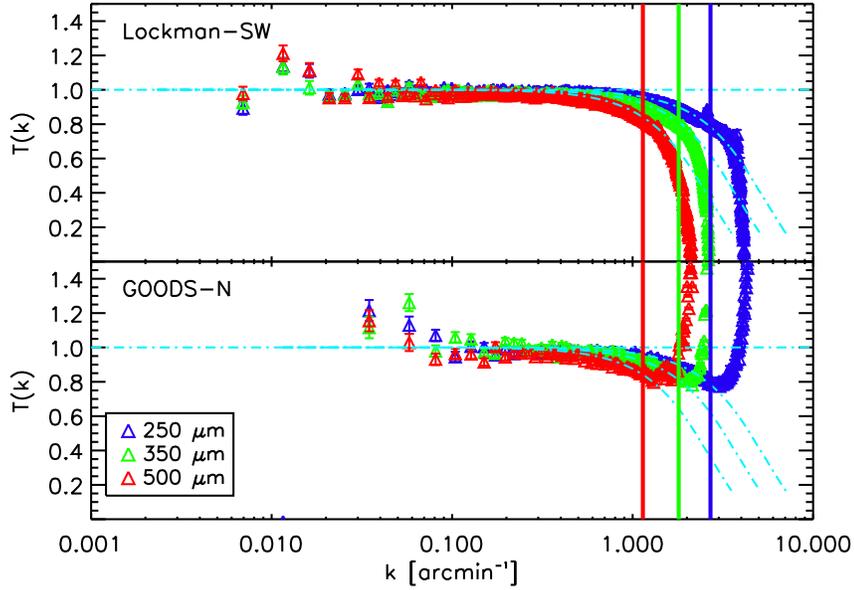}
}
\caption{{\bf Map-making transfer function.} Transfer function $T(k)$ due to the iterative map-maker and filtering on the cross-spectra of our two fields.
The blue, green and red triangles represent, respectively, the transfer function at 
250, 350 and 500 $\mu$m. $T(k)$ is essentially equal to 1 between 0.02 and 0.4 arcmin$^{-1}$.
The map-maker adds about 10\% power around 0.01 arcmin$^{-1}$ in the case of Lockman-SWIRE (top panel) and around 0.05 arcmin$^{-1}$ in the case
of GOODS-N (bottom panel) and reduces the power on small scales
mostly by averaging the data into pixels (light blue dotted dashed lines).
The vertical lines mark the maximum $k$ out to which we make use of the power spectrum estimates for shot-noise and
clustering measurements.
}
\label{transpk}
\end{figure}

To measure the transfer function $T(k)$ associated with the map maker,
we realize 100 simulations of a first estimate of our beam-convolved power spectrum and pass it through
the iterative map-making pipeline used to reduce our real data. We then compute the average of the ratio between
the estimated spectrum and the input spectrum with simulated maps masked exactly as in the real sky maps. 
This function is the transfer function $T(k)$ of the map-making pipeline associated with median
filter and other filtering (Figure~S\ref{transpk}). We divide the estimated power spectrum by this transfer function to remove the map making pipeline processing
effects.

\subsection{Total error budget:} 
\begin{figure}
\hspace{-0.7cm}
\centerline{
\includegraphics[width=14cm]{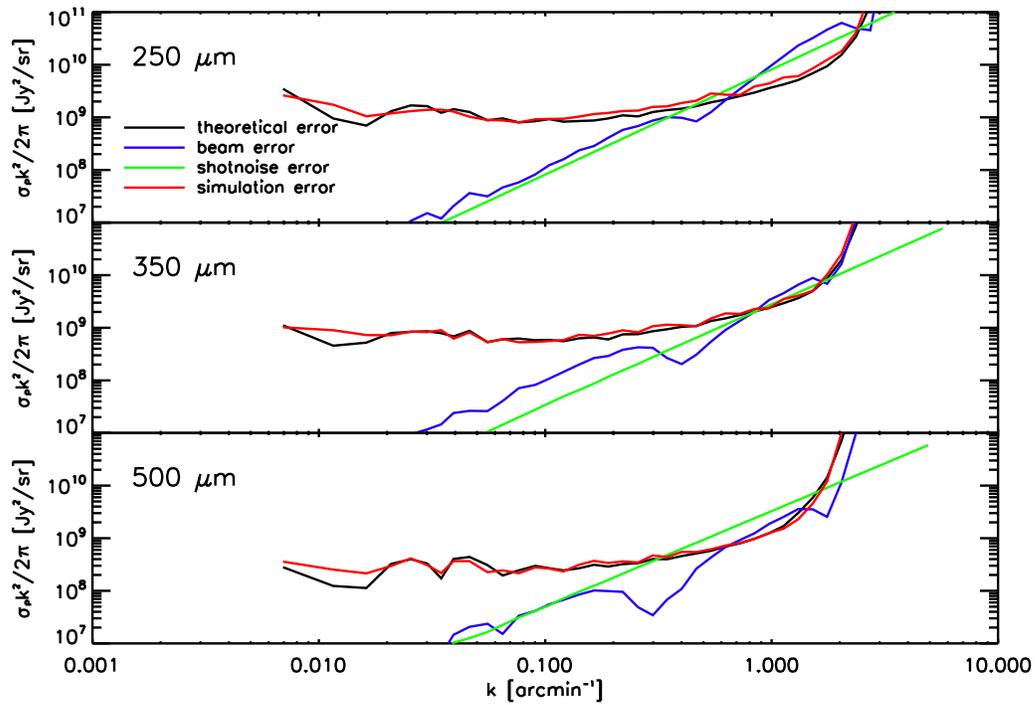}
}
\caption{{\bf The power spectrum uncertainties.} Error budget at 250, 350, and 500 $\mu$m.
We show the error separated into the beam uncertainty (blue lines), the shot-noise determination (green lines), and the simulations (sky and instrumental variance, red lines). The simulation uncertainty is compared to an analytical noise estimate (black lines).} 
\label{errorpk}
\end{figure}

The total error budget in our clustering plots is composed  of three contributions involving the uncertainty of the beam,  uncertainty in the shot-noise
determination, and the instrumental noise error. The latter is computed from simulations while the beam error comes from the differences in our estimates of the beam.
Shot-noise error results from direct fits to the measured data points. Figure~S\ref{errorpk} summarizes the error budget as a function of wave number and
also compares the instrumental noise error to an analytical formula for its expectation\cite{Knox95}.

\begin{figure}
\centerline{
\includegraphics[width=8cm]{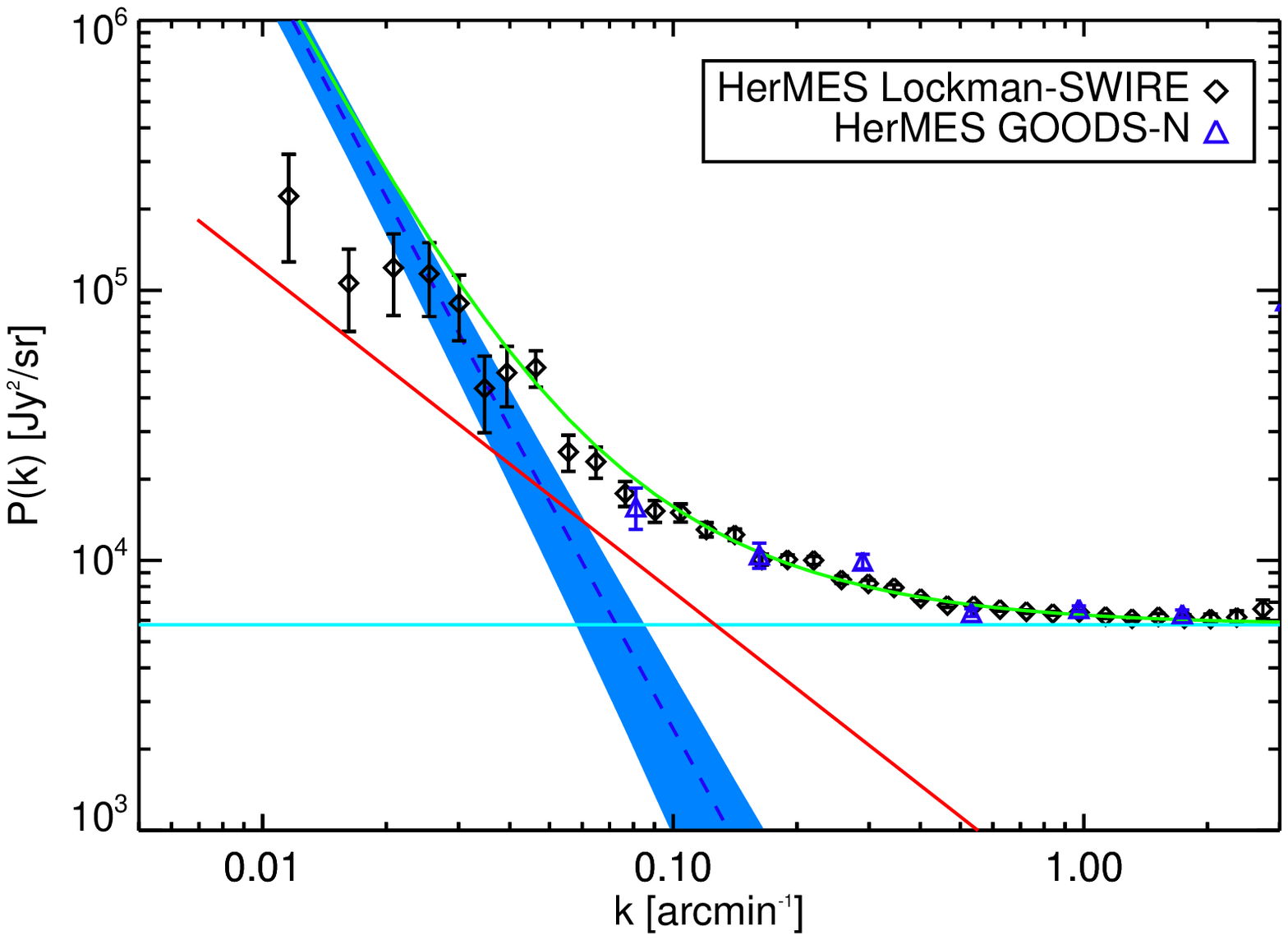}
\includegraphics[width=8cm]{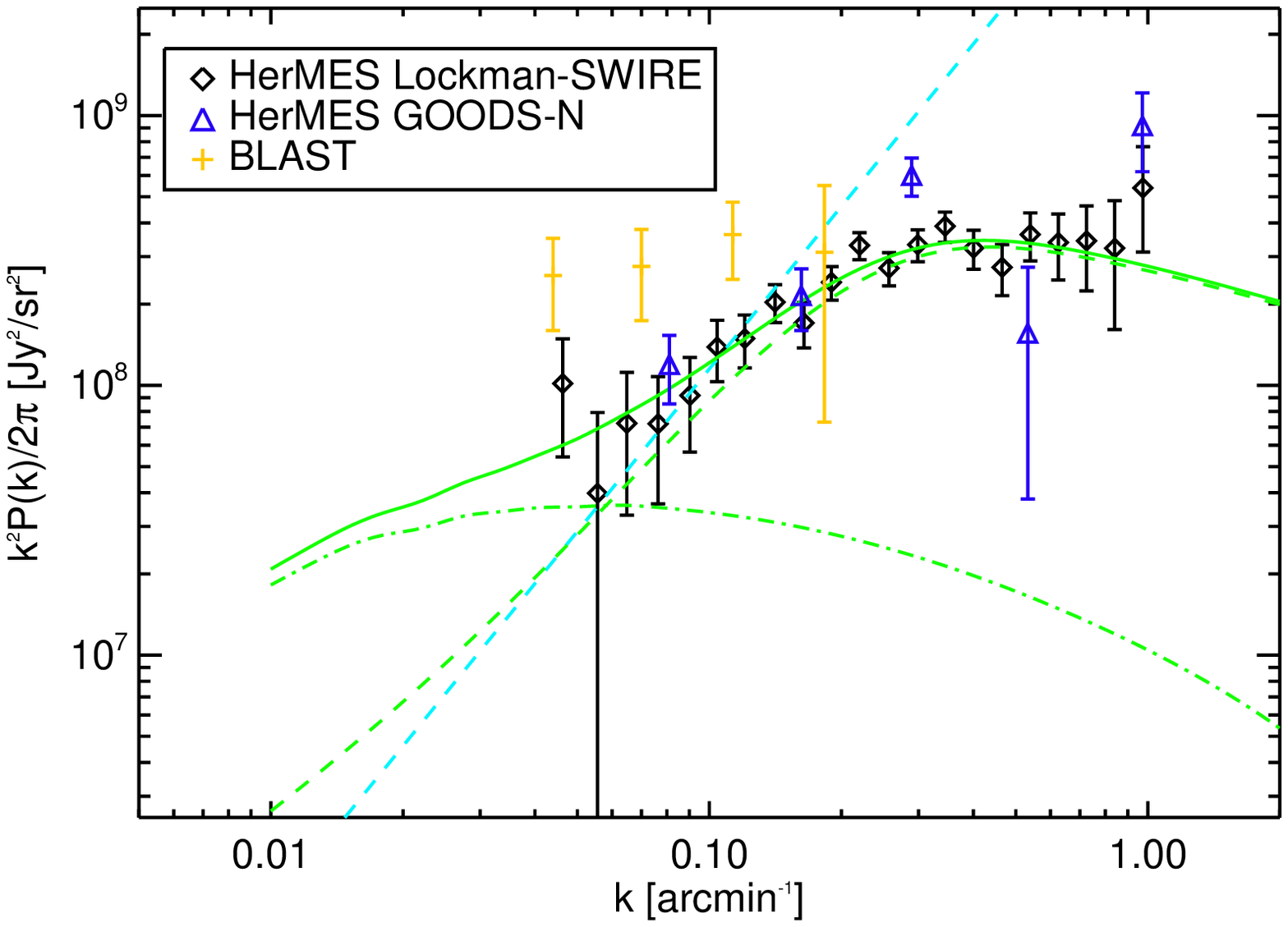}
}
\centerline{
\includegraphics[width=8cm]{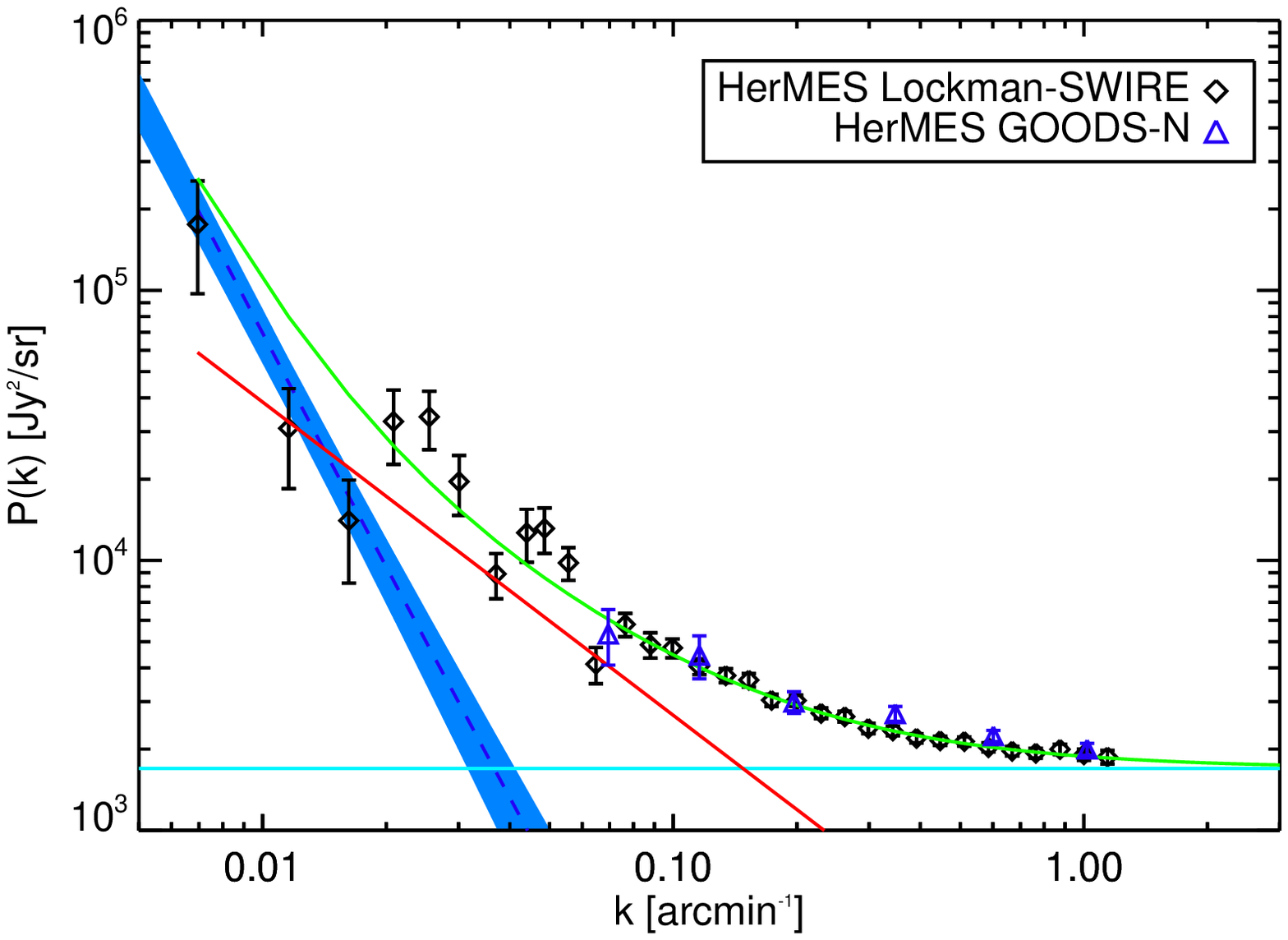}
\includegraphics[width=8cm]{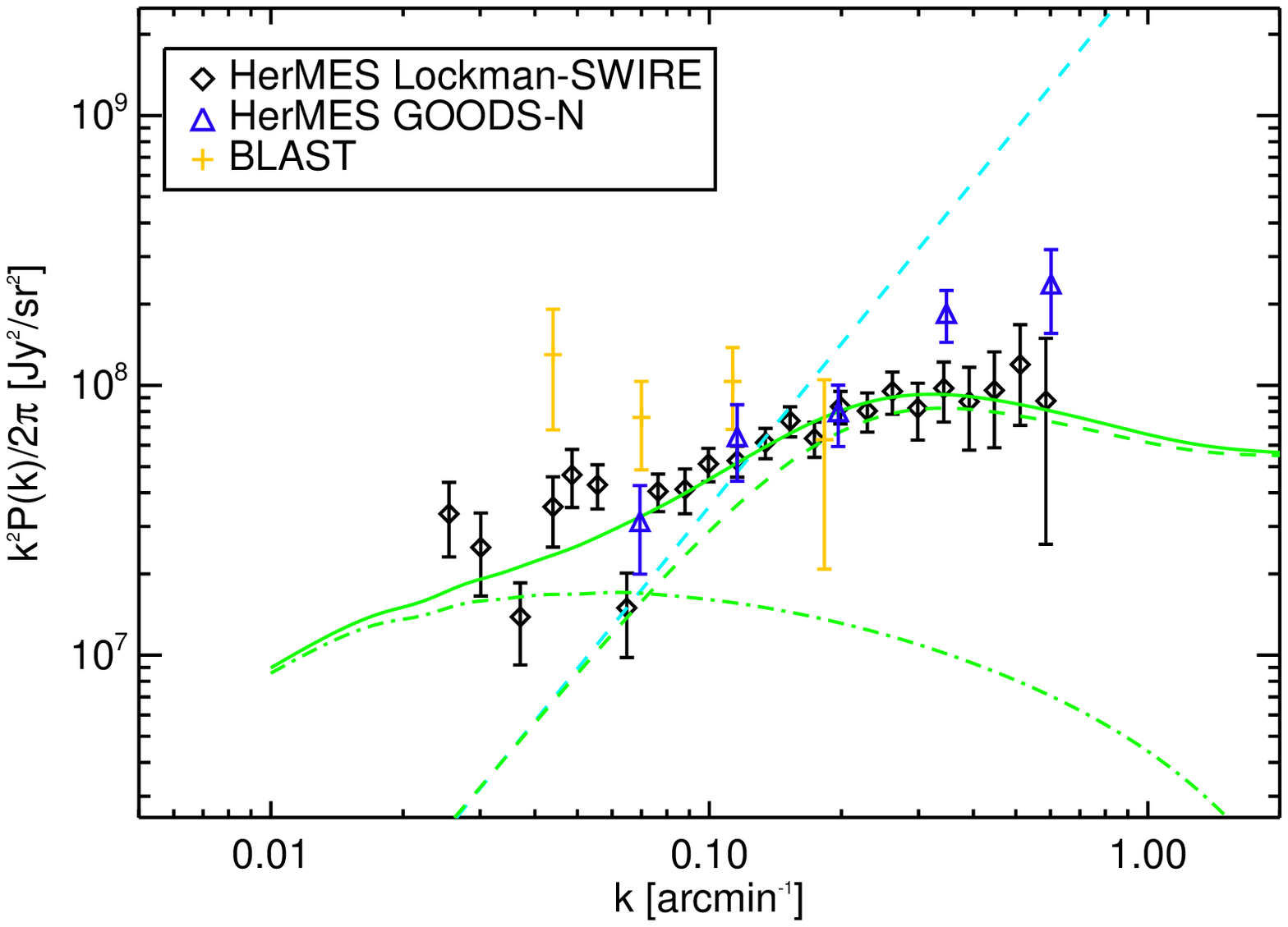}
}
\caption{{\bf The fluctuation power spectrum and the clustering component.}
The total power spectrum $P(k)$ (left panels) and clustering $P(k)$ with shot-noise removed (right panels) at 
250 $\mu$m (top) and 500 $\mu$m (bottom), respectively. 
The  power spectrum measurements shown are binned logarithmically for $k>0.03$ arcmin$^{-1}$, with a bin width equal to $\Delta k/k=0.15$, 
and linearly for smaller $k$, with a bin width of $\Delta k = 4.6\times10^{-3}$ arcmin$^{-1}$. 
This figure is similar to Fig~1.
}
\label{pkpowerlaw}
\end{figure}

While in Figure~1 of the main paper we showed $P(k)$ at 350 $\mu$m, in Figure~S\ref{pkpowerlaw} we show the same at 250 and 500 $\mu$m.

\subsection{Jack-knife tests:}

The results shown in this study were obtained by dividing the data into 
two equal and consecutive halves and by taking the cross-power spectrum of the
resultant maps. We can make the same cross-correlation power spectrum
measurements and repeat the whole process with maps made by dividing data 
into several other combinations. To be specific, we divide the data into four pieces, each filling
approximately our total field, and use these to measure the cross-power spectrum for two other combinations namely
$[(1+3) \times (2+4)]$ and $[(1+4) \times (2+3)]$, where 1 to 4 are four equal subdivisions of data in time.

\begin{figure}
\hspace{-0.7cm}
\centerline{
\includegraphics[width=14cm]{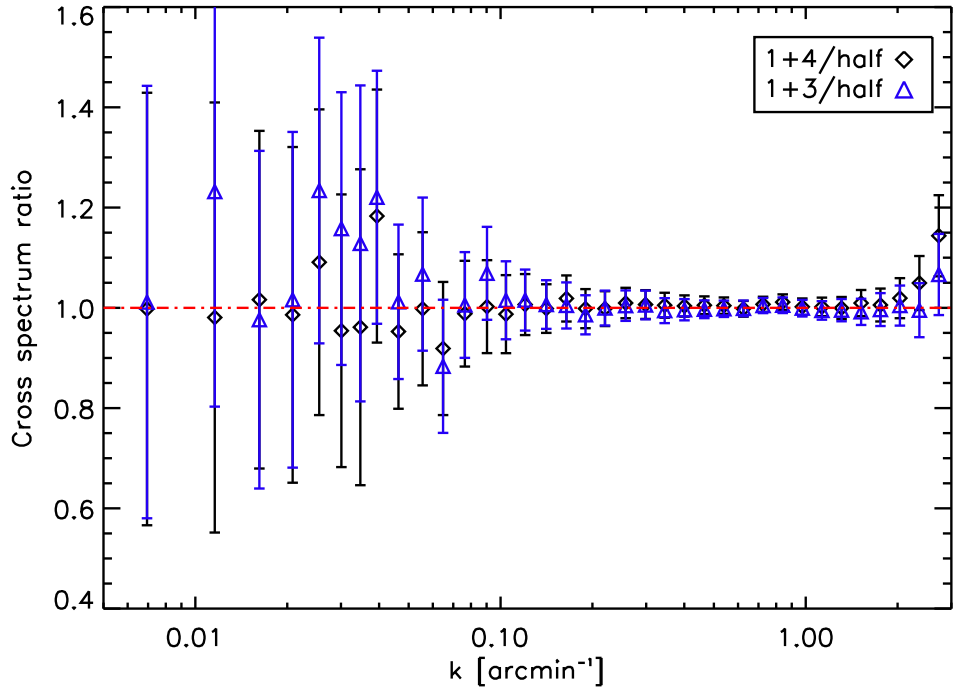}
}
\caption{{\bf Accuracy of the power spectrum measurement.} Ratios of the total power spectra $P(k)$ estimated with different sub-maps of Lockman-SWIRE data normalized to the default power
spectrum shown in the main paper estimated with the $(1+2)$ map cross-correlated against the $(3+4)$ map, after the data
are divided into four sequential intervals in time, labeled 1 to 4, of equal duration. The 1 and 3 subsets have the same scan direction
and the 2 and 4 subsets have the same scan directions, but the two subsets are almost orthogonal.
}
\label{jackknife}
\end{figure}

Figure~S\ref{jackknife} summarizes our results, showing that within the uncertainties 
we recover similar power spectra.
Given the observing strategy, the $(1+3)$ and $(2+4)$ maps are each made with parallel scans,
but roughly perpendicular to each other. The fact that we do not see a statistically significant 
difference shows that the beam ellipticity is not an important systematic concern in this study.

\subsection{Null tests:}

\begin{figure}
\hspace{-0.7cm}
\centerline{
\includegraphics[width=9cm]{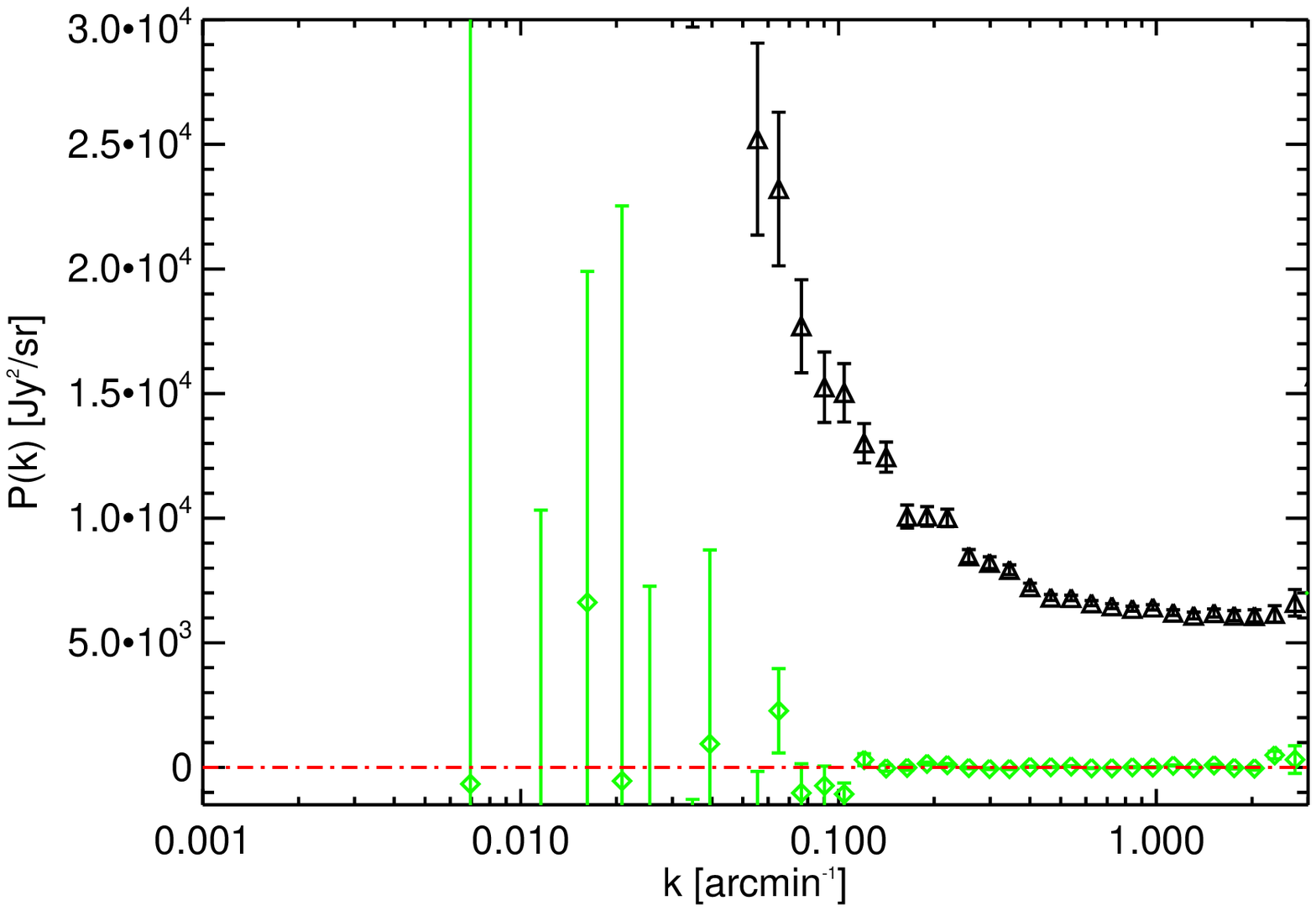}
\includegraphics[width=9cm]{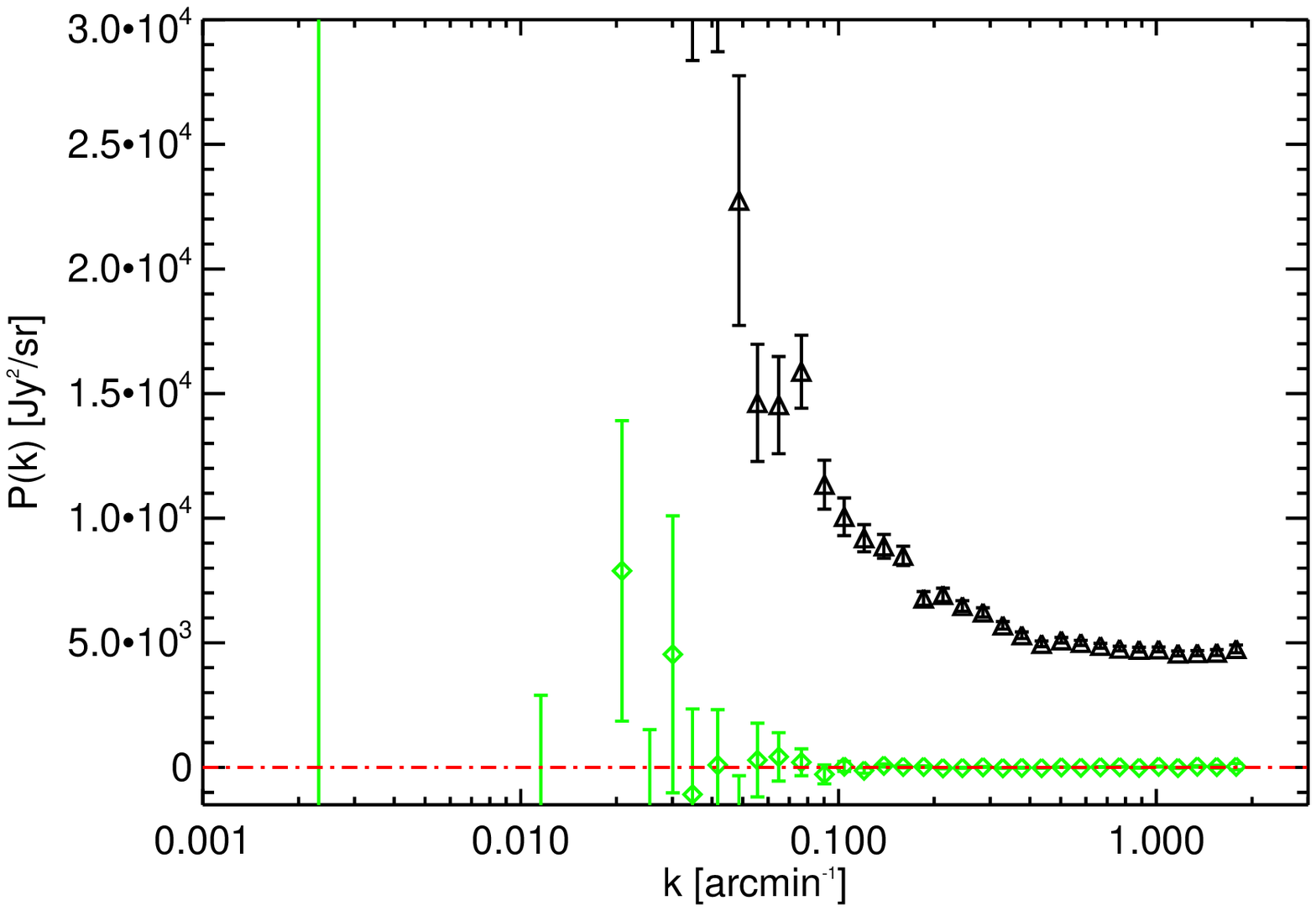}
}        
\centerline{
\includegraphics[width=9cm]{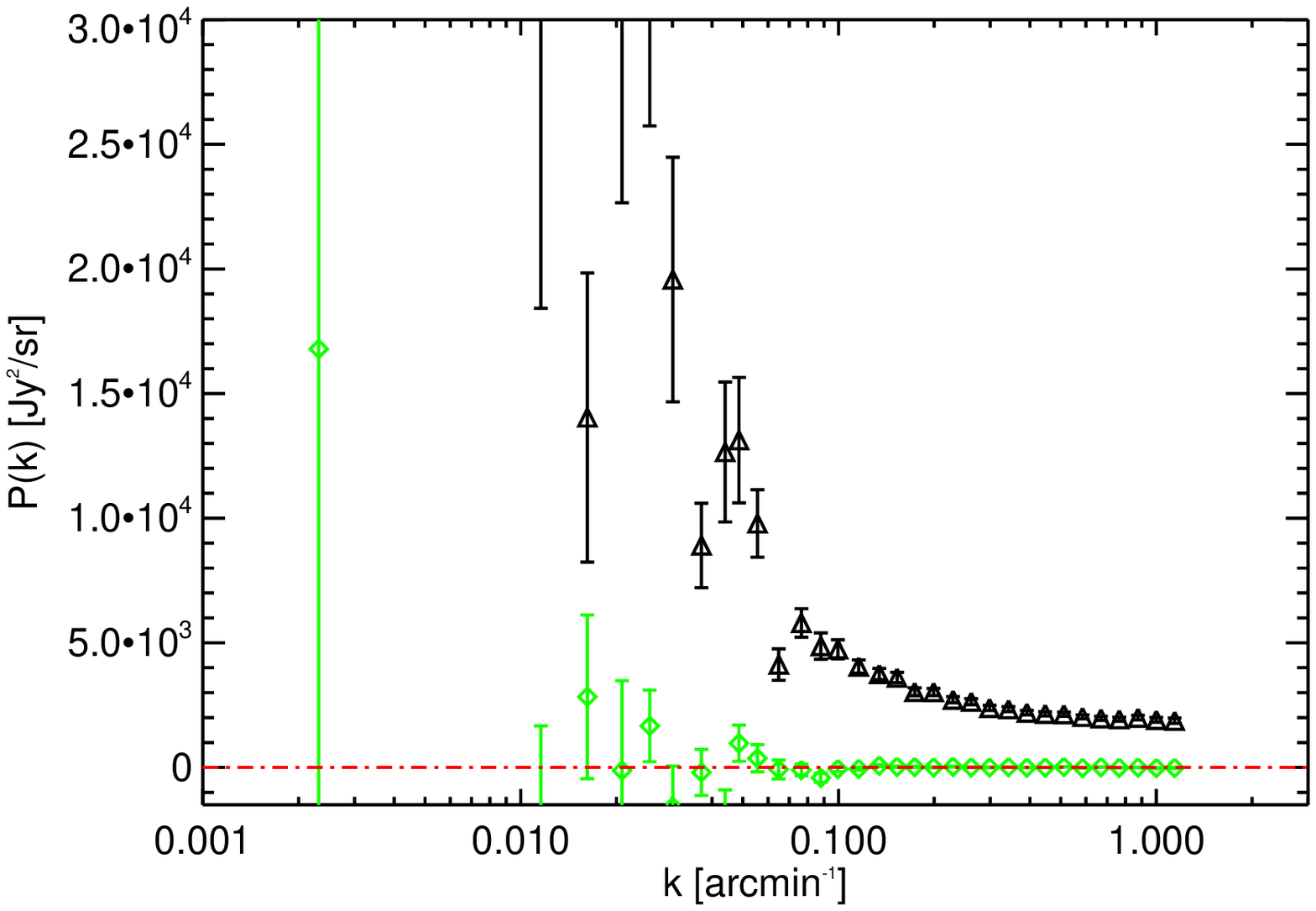}
}
\caption{{\bf The null-test of the power spectrum measurement.}
$P(k)$ measured (black triangle) on Lockman-SWIRE with the cross spectrum
  $[(1+2) \times (3+4)]$ at 250, 350, 500 $\mu$m (left to right, top
  to bottom). Cross power spectrum (green diamond) of the difference $[(1-2) \times (3-4)]$.
}
\label{nulltest}
\end{figure}

In addition to the jack-knife tests with a variety of sub-maps with data divided to four intervals and all leading to a measurement
of the sky signal,  we also perform several null tests using data combinations that remove the sky signal.
In this case, instead of computing the cross-power spectra of the sum maps of data combinations,
we make use of the sub-maps made by taking the differences of data combinations, again data divided to four sub-intervals
as the case of signal measurement. As an example, in Figure~S\ref{nulltest} we show the cross-power spectrum
computed at 250, 350, and 500 $\mu$m with the $(1-2)$ map cross-correlated against the $(3-4)$ map.  For reference,
we also show the default power spectrum computed with
$[(1+2) \times (3+4)]$.

\begin{table}
\begin{center}
\begin{tabular}{c c c c c}
\hline\hline
Band &  A  (Jy$^2$/sr)& $n$ & $P_{\rm SN}$ (Jy$^2$/sr) & $\chi^2$/d.o.f.\\ [0.5ex]
%heading
\hline
250 $\mu$m &  $(7.64 \pm 0.55$)$\times10^3$ & $-1.20 \pm0.09$ & $5798^{+92}_{-132}$ & $0.93$ \\
350 $\mu$m &  $(5.79 \pm 0.26$)$\times10^3$ & $-1.28 \pm0.07$ & $4373^{+62}_{-76}$ & $1.03$\\
500 $\mu$m &  $(2.67 \pm 0.13$)$\times10^3$ & $-1.16 \pm0.09$ & $1700 \pm 80$ & $1.2$\\ 
\hline
\end{tabular}
\caption{{\bf Power-law best fit values at $k_1=0.1$ arcmin$^{-1}$}. }
\end{center}
Notes: To describe the power spectrum, we take a power-law with
$P(k)=A(k/k_1)^n+P_{\rm SN}$ where $k_1$ is fixed at $0.1$
arcmin$^{-1}$ and $P_{\rm SN}$ is the shot-noise amplitude, assuming a power-law fit to the data. The errors are 68\% confidence level uncertainties determined
from the MCMC fits.
\label{table:powerlaw}
\end{table}

\subsection{Cirrus power spectrum}

The cirrus signal in our Lockman-SWIRE field is taken from existing measurements in the same field 
with {\it IRAS} 100 $\mu$m and MIPS\cite{Lagache07a}  with a power spectrum, $P(k)$, of the form $P_{\rm cirrus}(k)=P_0(k/k_0)^\beta$ at 160 $\mu$m 
with $P_0=(2.98 \pm 0.66)\times 10^6$ Jy$^2$/sr and $\beta=-2.89 \pm 0.22$ when $k_0=0.01$ arcmin$^{-1}$. We 
and extend this spectrum from 100 $\mu$m to SPIRE wavelengths  using the spectral dependence of a Galactic dust  model\cite{FDSa}.

\begin{table}
\begin{center}
\begin{tabular}{c c c c c c }
\hline\hline
 &   $ log[M_{\rm min}/M_{\odot}]$ &$\alpha$ & $\langle b \rangle_z$ & $P_{\rm SN}$ (Jy$^2$/sr) & $\chi^2$/d.o.f.\\ [0.5ex]
%heading
250 $\mu$m & $11.1^{+1.0}_{-0.6}$  & $1.6^{+0.1}_{-0.2}$ & $2.0^{+0.9}_{-0.1}$ & $6100 \pm 120$ & $0.76$\\
350 $\mu$m & $11.5^{+0.7}_{-0.2}$  & $1.8^{+0.1}_{-0.7}$ & $2.4^{+1.0}_{-0.2}$ & $4600 \pm 70$ & $1.02$ \\
500 $\mu$m & $11.8^{+0.4}_{-0.3}$  & $1.8^{+0.1}_{-0.7}$ & $2.8^{+0.4}_{-0.5}$ & $1800 \pm 80$ & $1.44$ \\
\hline
\end{tabular}
\caption{{\bf Halo model best fit values from the measured power spectra at the three wavebands.}}
\end{center}
\label{table:nonlin}
We tabulate the best-fit values with 68\% confidence level errors for halo occupation number used to interpret the power spectrum measurements.
The average galaxy bias factor is  $\langle b \rangle_z$. $P_{\rm SN}$ is the amplitude of shot-noise
fluctuations, also jointly determined from the power spectra as part of our model fitting process. 
The errors of the shot-noise amplitudes $P_{\rm SN}$ include an
extra error corresponding to the uncertainty of the absolute flux calibration scale at the three SPIRE wavebands of 15\%\cite{Swinyard2010a}.
The chi-square values of the best-fit model, per degree-of-freedom, are also tabulated.
We do not tabulate the values of $M_1$ as it remains unconstrained within the prior of $M_1/M_{\rm min}$  taken to be between 10 and 25.
\end{table}

\clearpage
\newpage

\subsection{Interpretation Model}

In Table~S1 we summarize results related to power-law model fits with $P(k)=A(k/k_1)^n+P_{\rm SN}$,
where $k_1= 0.1$ arcmin$^{-1}$.  We also make use of the halo model  approach to model-fit
our clustering measurements, making use of the halo occupation distribution (HOD)\cite{Cooray2002a}.

The number of pairs of galaxies inside a given halo depends on the 
variance of the HOD, $\sigma^2(M,z)=\langle\,N_{\rm gal}(N_{\rm gal}-1)\rangle$
while the number of pairs of galaxies in different halos is simply given 
by the square of the mean halo occupation with
$N(M,z)=\langle\,N_{\rm gal}\rangle$, where $N_{\rm gal}$
is the total number of galaxies in a halo and we further assume 
that one galaxy occupies the center of the halo, the others being 
considered as satellite galaxies, so that $N_{\rm gal}=N_{\rm cen}+N_{\rm sat}$. 
Central and satellite galaxies are assumed to have different HODs; in fact the mean number 
of central galaxies in a given halo is a simple step function, so that $N_{\rm cen}=1$ 
above a given mass $M_{\rm min}$, and $N_{\rm cen}=0$ otherwise. The HOD of satellite galaxies 
is taken to be a power law of the halo mass\cite{Zehavi2005}:
\begin{equation}
N_{\rm sat}=\left(\frac{M}{M_1}\right)^{\alpha};
\end{equation}
here $M_1$ is a normalization factor that represents 
the mass scale at which a single halo hosts on average one satellite galaxy in addition to the central 
galaxy.

The power spectrum of galaxies 
is then parameterized as the sum of two different contributions:
the 1-halo term, which describes the clustering on small scales and is related 
to pairs of galaxies within the same 
halo and the 2-halo term, responsible for the large scale clustering 
and related to pairs of galaxies in different halos:
\begin{equation}
P(k,z)=P_{\rm 1h}(k,z)+P_{\rm 2h}(k,z).
\end{equation}
The two terms are then written as:
\small
\begin{eqnarray}P_{\rm 1h}(k,z) & = & \int dM\, \frac{dn_{\rm 
halo}}{dM}(z)[2N_{\rm cen}(M)N_{\rm sat}(M) u_{\rm DM}(k,z|M) + \nonumber 
\\
& &  ~~~~~~ N^2 _{\rm sat}(M) u^2 _{\rm DM}(k,z|M) ] dM/n_{\rm 
gal}^{2}(z),\nonumber \\
%\label{eq:P1h}\end{eqnarray}
%\normalsize
%\newline
& & \nonumber \\
%\small
%\begin{eqnarray}
  P_{\rm 2h}(k,z) & = & P_{\rm DM}(k,z)\times
 \nonumber \\ 
  & &  ~~~~~~  \Big{[} \int dM\, \frac{dn_{\rm halo}}{dM}(z)N_{\rm gal}(M,z)\times 
\nonumber \\
  & &   ~~~~~~ b(M,z)u_{\rm DM}(k,z|M)dM\Big{]}^{2}/n_{\rm gal}^2(z). \nonumber 
\normalsize
\\
  \label{eq:P2h}
\normalsize
\end{eqnarray}
\normalsize
Here $P_{\rm DM}(k,z)$ is the linear 
dark matter power spectrum; $n_{\rm halo}$ is the halo-mass function\cite{Sheth2001}; 
$b(M,z)$ is the linear bias which connects the large scale clustering of dark matter to the 
galaxy clustering; $u_{\rm DM}(k,z|M)$ is the normalized dark matter halo density 
profile in Fourier space (as a function of wavenumber $k$ and redshift $z$ for a given value of mass
 $M$) and $n_{\rm gal}$ is the mean number of galaxies per unit comoving volume at redshift $z$:
\begin{equation}
  n_{\rm gal}(z) = \int dM\, \frac{dn_{\rm halo}}{dM}(z)\Bigg{[} 1 + \left( 
\frac{M}{M_1} \right) ^{\alpha}\Bigg{]}.
\normalsize
\end{equation}
For the dark matter halo density function we adopt the Navarro-Frenk-White\cite{NFW} 
profile  truncated at the virial radius $r_{\rm vir}$ and with a 
concentration parameter given by:
\begin{equation}
c(M,z)=\frac{9}{1+z}\left(\frac{M}{M_{*}}\right)^{-0.13};
\end{equation}
 here $M_{*}$ is the characteristic mass scale at which the critical density required
 for spherical collapse is equal to the square root of the variance in the initial density 
field  $\sigma(M)$ when extrapolated at the present time using linear theory such that 
$\delta_{\rm sc}(z)=\sigma(M_*)$, where $\delta_{\rm sc}(z)=1.68/g(z)$, where $g(z)$ is the linear theory
growth function for density perturbations.

As outlined above $M_{\rm min}$ determines both the one-halo and two-halo amplitudes, while
 $\alpha$  determines primarily the amplitude of the one-halo term and the overall number density of galaxies, which in return is
connected to the amplitude of the two halo term via the halo bias factor. While with the two-halo term alone all these parameters are degenrate with each 
other and the bias factor, allowing only an average mass scale to be determined based on the bias factor, 
with one-halo term also included some of the degeneracies are broken and $M_{\rm min}$ and $\alpha$ 
can  be determined independent of bias\cite{Cooray2002}.

\begin{figure}
\hspace{-0.7cm}
%\hspace{-2cm}
\centerline{
 \includegraphics[width=18cm]{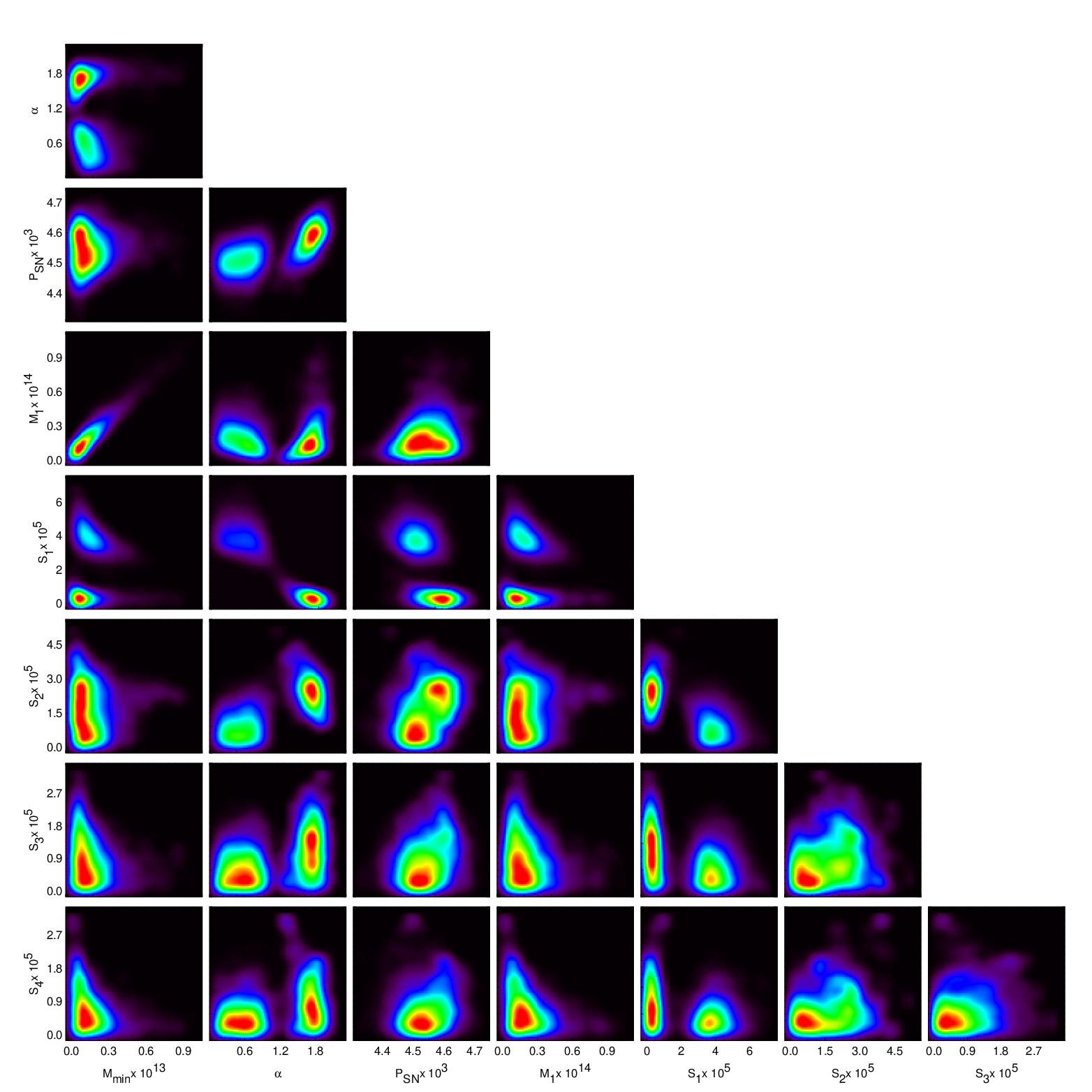}
}
\caption{{\bf The halo model parameter estimates.} Bi-dimensional probability distribution function for all the
pairs  associated with our halo model fits with eight parameters ($M_1, M_{min}, \alpha,
  P_{SN}, S_1, S_2, S_3$ and $S_4$) showing our constraints and the
  degeneracies between the parameters. Here we show results at 350 $\mu$m, but degeneracies of parameters related to 250 and 500 $\mu$m model
fits are similar.
}
\label{triangle}
\end{figure}

In the Limber approximation, the measured power spectrum of fluctuations can be expressed 
as the 2 dimensional, flux 
averaged projection of the three-dimensional galaxy power spectrum $P(k,z)$ as:
\begin{equation}
P(k_{\theta})=\int_{z_{\rm min}}^{z_{\rm max}}P\left(k=\frac{2\pi\,k_{\theta}}{x(z)},z\right)\left(\frac{d{\cal S}}{dz}(z)\right)^2\frac{1}{dV_{\rm c}(z)}dz;
\end{equation} 
here $d{\cal S}/dz$ is the redshift distribution of the cumulative flux contributed by the background faint galaxies, 
$dV_{\rm c}$ is the comoving volume element, defined as $dV_{\rm c}\equiv x(z)^2\frac{dx}{dz}$ and $x(z)$ 
is the comoving radial distance. 

In this paper  we determine the redshift distribution of the intensity by
binning the redshift range in four redshift bins between $z=0$ and $z=4$ and putting constraints on $d{\cal S}/dz$ 
in each bin; the advantage of this approach is that we don't assume a particular model for $d{\cal S}/dz(z)$; 
instead, we let the data decide which model is more adequate.

The method we use to constrain our parameters is based on a modified
version of the publicly available Markov-Chain Monte-Carlo (MCMC) package CosmoMC\cite{Lewis2002}, 
with a convergence diagnostics based on the Gelman-Rubin
criterion\cite{gelman}. We consider a halo model described by the
following set of parameters:
\begin{equation}
 \label{parameter}
      \{dS/dz_i,M_{\rm min},\alpha,M_1,P_{\rm SN}\}~,
\end{equation}
where, as discussed before, we bin the cumulative flux $dS/dz(z)$ in four redshift
bins, $dS/dz_i(z)\,(i=1,2,..4)$, representing the value at four redshift intervals, 
${\rm bin}_i\in{[0-1,1-2,2-3,3-4]}$. 
In the above $P_{\rm SN}$ is the shot-noise amplitude which we remeasure again for the halo model fits.
To obtain reliable model-fits to data we set a broad uniform prior on the ratio $M_1/M_{\rm min}$ to be
between 10 to 25, consistent with numerical simulations of the halo occupation distribution which
finds a value close to 15 for this ratio\cite{Zheng2005}. 
We also require that the redshift integrated source intensity be within the 68\% confidence level ranges of
the  background light intensities as obtained by FIRAS\cite{Fixsen1998a}.
The central values and errors we use are $0.85 \pm 0.08$, $0.65 \pm 0.19$ and $0.39 \pm 0.10$ MJy/sr
at 250, 350, and 500 $\mu$m, respectively. For background cosmology, we assume the concordance model\cite{Komatsu2010}.
Our results related to the halo model fits are summarized in Table~S2. 

\begin{figure}
\hspace{-0.7cm}
%\hspace{-2cm}
\centerline{
 \includegraphics[width=14cm]{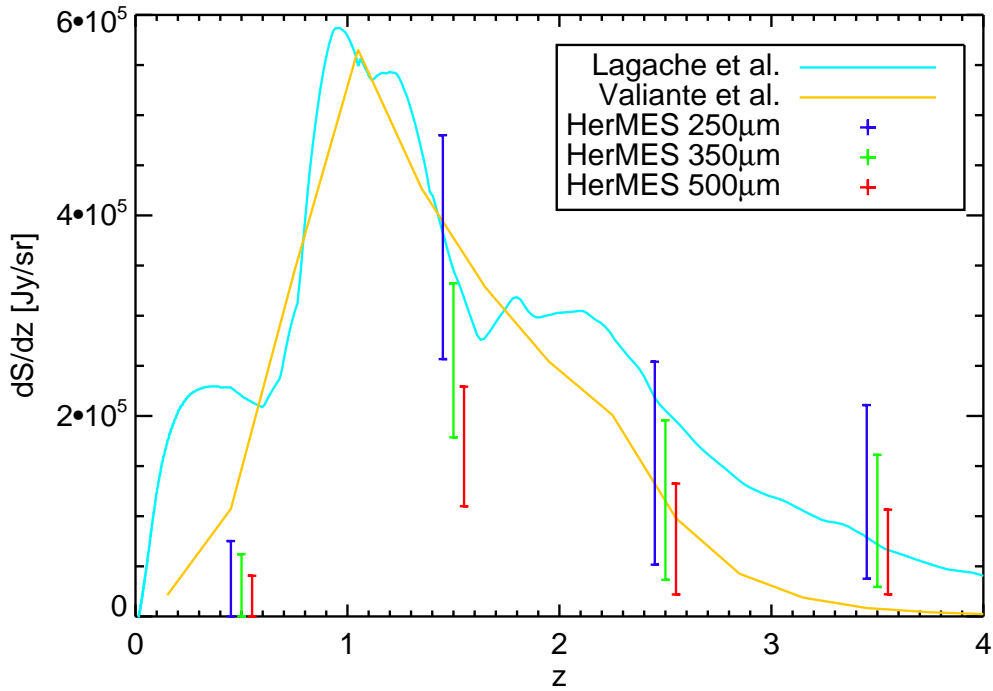}
}
\caption{{\bf Redshift evolution of the galaxy intensities at the sub-millimetre wavelengths.} 
$d{\cal S}/dz$ as a function of redshift for 4 bins in redshift and for the three wavebands of SPIRE with $S < 50$ mJy. 
For reference, we show 2 model predictions from Lagache et al.\cite{Lagache2003a} and Valiante et al.\cite{Valiante2009a}
for the same flux cut. We have used a prior on the occupation number slope $\alpha > 1$.
}
\label{dsdz}
\end{figure}

In comparison to the shot-noise values from model fits to the power spectrum (Table~S1 for the power-law case and Table~S2 for the halo model case),
 the shot-noise values from the best determined source counts\cite{Glenn2010a} give
$6900 \pm 320$, $4500 \pm 220$, and $1600 \pm 100$  Jy$^2$/sr at 250, 350, and 500 $\mu$m, respectively.

In Figure~S\ref{triangle} 
we show the two-dimensional constraints on pairs of parameters
that highlight the degeneracies associated with this eight parameter model fit.
The best-fit values and the errors at each of the three wavebands are show in
Figure~S\ref{dsdz}.

Additionally, we compute the far-infrared bolometric luminosity between 8 and 1100 $\mu$m in each of the redshift bins from the $dS/dz(z)$ values
by modelling the flux received between a redshift $z^i_{\rm min}$ and $z^i_{\rm max}$ in each
$j$ SPIRE bands, defined by the bandpass $f_j(\nu)$:
\begin{equation}
  \label{luminosity}
  dS^j/dz^i=L_{\rm FIR}\int_{z^i_{\rm min}}^{z^i_{\rm
      max}}dz{(1+z)^{(\beta+1)}\over4\pi
    D_L^2(z)}\frac{\displaystyle \int\nu^\beta
      B(\nu(1+z),T)f_j(\nu)d\nu}{\displaystyle \int_{270\, \rm GHz}^{38\, \rm THz}\nu^\beta B(\nu,T)d\nu}
\end{equation}
The temperature T is chosen to be $28\pm8$ K and the emissivity index $\beta$ is fixed to 1.5,
we then fit for $L_{\rm FIR}$ given the measured values and the predicted values of $dS/dz_i$.
The temperature uncertainty is incorporated into the $L_{\rm FIR}$ error budget.
We summarize our results related to $L_{\rm FIR}$ as a function of redshift in Figure~2. $L_{\rm FIR}$
is a measure of the star-formation rate with\cite{Kennicutt1998a}
\begin{equation}
{\rm SFR}[{\rm M}_{\odot} {\rm yr}^{-1}] = 1.73\times10^{-10}\, L [{\rm L}_{\odot}] \, .
\end{equation}
We use this to also show the SFR implied by $L_{\rm FIR}$  in Figure~2. 
Here we have subselected the models  that lead to $\alpha >1$
to be consistent with the occupation numbers at other wavelenegths\cite{Cooray2002a,Zehavi2005}. 
$L_{\rm FIR}$ as a function of redshift has been predicted in two analytical models of
sub-millimetre galaxy population\cite{Lagache2003a,Valiante2009a} and we make a comparison in the same figure.

\end{document}